\documentclass[twocolumn]{aastex62}
\usepackage{natbib}
\usepackage[style=american]{csquotes}
\usepackage{epsfig}
\usepackage{apjfonts}
\usepackage{lmodern} 
\usepackage{graphicx}
\usepackage{tabularx}   

\providecommand{\msun}{\ensuremath{\,M_\Sun}}
\providecommand{\rsun}{\ensuremath{\,R_\Sun}}
\providecommand{\lsun}{\ensuremath{\,L_\Sun}}
\providecommand{\mj}{\ensuremath{\,M_{\rm J}}}

\providecommand{\me}{\ensuremath{\,M_{\rm E}}}
\providecommand{\re}{\ensuremath{\,R_{\rm E}}}

\newcommand{\tess}{TESS}

\newcommand{\mps}{m s$^{-1}$}

\accepted{22 June, 2020}

\shorttitle{}
\shortauthors{}

\begin{document}

\title{TESS Reveals a Short-period Sub-Neptune Sibling (HD 86226c) to a Known Long-period Giant Planet \footnote{This paper includes data gathered with the 6.5 m Magellan Telescopes located at Las Campanas Observatory, Chile.}}

\correspondingauthor{Johanna Teske}
\email{jteske@carnegiescience.edu}

\author{Johanna Teske}
\altaffiliation{NASA Hubble Fellow}
\affiliation{The Observatories of the Carnegie Institution for Science, 813 Santa Barbara Street, Pasadena, CA 91101, USA}

\author[0000-0002-2100-3257]{Mat\'ias R. D\'iaz}
\affiliation{Departamento de Astronom\'ia, Universidad de Chile, Camino El Observatorio 1515, Las Condes, Santiago, Chile}

\author{Rafael Luque}
\affiliation{Instituto de Astrof\'isica de Canarias (IAC), E-38205 La Laguna, Tenerife, Spain}
\affiliation{Departamento de Astrof\'sica, Universidad de La Laguna (ULL), E-38206, La Laguna, Tenerife, Spain}

\author{Teo Mo\v{c}nik}
\affiliation{Department of Earth and Planetary Sciences, UC Riverside, 900 University Ave., Riverside, CA 92521, USA}

\author[0000-0002-7990-9596]{Julia V. Seidel} 
\affiliation{Observatoire de l'Universit{\'e} de Gen{\`e}ve, 51 chemin des Maillettes, 1290 Versoix, Switzerland}

\author{Jon Fern\'andez Otegi}
\affiliation{Observatoire de l'Universit{\'e} de Gen{\`e}ve, 51 chemin des Maillettes, 1290 Versoix, Switzerland}
\affiliation{Institute for Computational Science, University of Zurich,  Winterthurerstr. 190, CH-8057 Zurich, Switzerland} 

\author[0000-0001-6039-0555]{Fabo Feng}
\affiliation{Earth \& Planets Laboratory, Carnegie Institution for Science, 5241 Broad Branch Road, NW, Washington, DC 20015, USA}

\author[0000-0003-2733-8725]{James S. Jenkins}
\affiliation{Departamento de Astronom\'ia, Universidad de Chile, Camino El Observatorio 1515, Las Condes, Santiago, Chile}

\author[0000-0003-0987-1593]{Enric Pall\`e}
\affiliation{Instituto de Astrof\'isica de Canarias (IAC), E-38205 La Laguna, Tenerife, Spain}
\affiliation{Departamento de Astrof\'sica, Universidad de La Laguna (ULL), E-38206, La Laguna, Tenerife, Spain}

\author{Damien S\'egransan} 
\affiliation{Observatoire de Gen\`eve, Universit\'e de Gen\`eve, 51 ch. des Maillettes, 1290 Sauverny, Switzerland}

\author[0000-0001-7576-6236]{St\`ephane Udry}
\affiliation{Observatoire de Gen\`eve, Universit\'e de Gen\`eve, 51 ch. des Maillettes, 1290 Sauverny, Switzerland}

\author[0000-0001-6588-9574]{Karen A.\ Collins}
\affiliation{Center for Astrophysics ${\rm \mid}$ Harvard {\rm \&} Smithsonian, 60 Garden Street, Cambridge, MA 02138, USA}

\author[0000-0003-3773-5142]{Jason D. Eastman}
\affiliation{Center for Astrophysics ${\rm \mid}$ Harvard {\rm \&} Smithsonian, 60 Garden Street, Cambridge, MA 02138, USA}

\author[0000-0003-2058-6662]{George R. Ricker}
\affiliation{Kavli Institute for Astrophysics and Space Research, Massachusetts Institute of Technology, Cambridge, MA 02139, USA}

\author[0000-0001-6763-6562]{Roland Vanderspek}
\affiliation{Kavli Institute for Astrophysics and Space Research, Massachusetts Institute of Technology, Cambridge, MA 02139, USA}

\author[0000-0001-9911-7388]{David W. Latham}
\affiliation{Center for Astrophysics ${\rm \mid}$ Harvard {\rm \&} Smithsonian, 60 Garden Street, Cambridge, MA 02138, USA}

\author[0000-0002-6892-6948]{Sara Seager}
\affiliation{Department of Physics and Kavli Institute for Astrophysics and Space Research, Massachusetts Institute of Technology, Cambridge, MA 02139, USA}
\affiliation{Department of Earth, Atmospheric and Planetary Sciences, Massachusetts Institute of Technology, Cambridge, MA 02139, USA}
\affiliation{Department of Aeronautics and Astronautics, MIT, 77 Massachusetts Avenue, Cambridge, MA 02139, USA}

\author[0000-0002-4265-047X]{Joshua N. Winn}
\affiliation{Department of Astrophysical Sciences, Princeton University, Princeton, NJ 08544, USA}

\author[0000-0002-4715-9460]{Jon M. Jenkins}
\affiliation{NASA  Ames  Research  Center,  Moffett  Field,  CA  94035, USA }

\author[0000-0001-7416-7522]{David. R. Anderson}
\affiliation{Astrophysics Group, Keele University, Staffordshire ST5 5BG, UK}
\affiliation{Department of Physics, University of Warwick, Gibbet Hill Road, Coventry CV4 7AL, UK}

\author[0000-0001-7139-2724]{Thomas Barclay}
\affiliation{NASA Goddard Space Flight Center, Greenbelt, MD 20771, USA}
\affiliation{University of Maryland, Baltimore County, 1000 Hilltop Circle, Baltimore, MD 21250, USA}

\author{Fran\c{c}ois Bouchy}
\affiliation{Observatoire de Gen\`eve, Universit\'e de Gen\`eve, 51 ch. des Maillettes, 1290 Sauverny, Switzerland}

\author[0000-0002-0040-6815]{Jennifer A. Burt}
\affiliation{Jet Propulsion Laboratory, California Institute of Technology, 4800 Oak Grove Drive, Pasadena, CA 91109, USA}

\author[0000-0003-1305-3761]{R. Paul Butler}
\affiliation{Earth \& Planets Laboratory, Carnegie Institution for Science, 5241 Broad Branch Road, NW, Washington, DC 20015, USA}

\author[0000-0003-1963-9616]{Douglas A. Caldwell}
\affiliation{NASA  Ames  Research  Center,  Moffett  Field,  CA  94035, USA }
\affiliation{Carl Sagan Center, SETI Institute, Moffett Field, CA, United States}

\author[0000-0003-2781-3207]{Kevin I.\ Collins}
\affiliation{George Mason University, 4400 University Drive, Fairfax, VA, 22030 USA}

\author[0000-0002-5226-787X]{Jeffrey D. Crane}
\affiliation{The Observatories of the Carnegie Institution for Science, 813 Santa Barbara Street, Pasadena, CA 91101, USA}

\author[0000-0001-6110-4610]{Caroline Dorn}
\affiliation{Institute for Computational Science, University of Zurich,  Winterthurerstr. 190, CH-8057 Zurich, Switzerland}

\author[0000-0001-8045-1765]{Erin Flowers}
\thanks{NSF GRFP Fellow}
\affiliation{Department of Astrophysical Sciences, Princeton University, Princeton, NJ 08544, USA}

\author[0000-0003-1231-2389]{Jonas Haldemann}
\affiliation{Department of Space Research \& Planetary Sciences, University of Bern, Gesellschaftsstrasse 6, CH-3012 Bern, Switzerland}

\author[0000-0001-5555-2652]{Ravit Helled}
\affiliation{Institute for Computational Science, University of Zurich,  Winterthurerstr. 190, CH-8057 Zurich, Switzerland}

\author[0000-0002-3439-1439]{Coel Hellier}
\affiliation{Astrophysics Group, Keele University, Staffordshire ST5 5BG, UK}

\author[0000-0002-4625-7333]{Eric L. N. Jensen}
\affiliation{Dept.\ of Physics \& Astronomy, Swarthmore College, Swarthmore PA 19081, USA}

\author[0000-0002-7084-0529]{Stephen R. Kane} 
\affiliation{Department of Earth and Planetary Sciences, UC Riverside, 900 University Ave., Riverside, CA 92521, USA}

\author[0000-0001-9380-6457]{Nicholas Law}
\affiliation{Department of Physics and Astronomy, University of North Carolina at Chapel Hill, Chapel Hill, NC 27599-3255, USA}

\author[0000-0001-6513-1659]{Jack J. Lissauer}
\affiliation{NASA  Ames  Research  Center,  Moffett  Field,  CA  94035, USA }

\author[0000-0003-3654-1602]{Andrew W. Mann}
\affiliation{Department of Physics and Astronomy, University of North Carolina at Chapel Hill, Chapel Hill, NC 27599, USA} 

\author[0000-0001-5630-1396]{Maxime Marmier}
\affiliation{Observatoire de l'Universit{\'e} de Gen{\`e}ve, 51 chemin des Maillettes, 1290 Versoix, Switzerland}

\author[0000-0002-5254-2499]{Louise Dyregaard Nielsen} 
\affiliation{Observatoire de l'Universit{\'e} de Gen{\`e}ve, 51 chemin des Maillettes, 1290 Versoix, Switzerland}

\author[0000-0003-4724-745X]{Mark E. Rose}
\affiliation{NASA  Ames  Research  Center,  Moffett  Field,  CA  94035, USA }

\author{Stephen A. Shectman}
\affiliation{The Observatories of the Carnegie Institution for Science, 813 Santa Barbara Street, Pasadena, CA 91101, USA}

\author[0000-0002-1836-3120]{Avi Shporer}
\affiliation{Department of Physics and Kavli Institute for Astrophysics and Space Research, Massachusetts Institute of Technology, Cambridge, MA 02139, USA}

\author[0000-0002-5286-0251]{Guillermo Torres}
\affiliation{Center for Astrophysics ${\rm \mid}$ Harvard {\rm \&} Smithsonian, 60 Garden Street, Cambridge, MA 02138, USA}

\author{Sharon X. Wang}
\affiliation{The Observatories of the Carnegie Institution for Science, 813 Santa Barbara Street, Pasadena, CA 91101, USA}

\author[0000-0003-2862-6278]{Angie Wolfgang}
\affiliation{Department of Astronomy \& Astrophysics, The Pennsylvania State University, 525 Davey Laboratory, University Park, PA 16802, USA}
\affiliation{Center for Exoplanets and Habitable Worlds, The Pennsylvania State University, 525 Davey Laboratory, University Park, PA 16802, USA}

\author[0000-0001-9665-8429]{Ian Wong}
\affiliation{Department of Earth, Atmospheric and Planetary Sciences, Massachusetts Institute of Technology, Cambridge, MA 02139, USA}

\author[0000-0002-0619-7639]{Carl Ziegler}
\affiliation{Dunlap Institute for Astronomy and Astrophysics, University of Toronto, Ontario M5S 3H4, Canada}

\begin{abstract}
The Transiting Exoplanet Survey Satellite mission was designed to find transiting planets around bright, nearby stars. Here we present the detection and mass measurement of a small, short-period ($\approx\,4$\,days) transiting planet around the bright ($V=7.9$), solar-type star HD 86226 (TOI-652, TIC 22221375), previously known to host a long-period ($\sim$1600 days) giant planet. HD 86226c (TOI-652.01) has a radius of $2.16\pm0.08$~$R_{\oplus}$ and a mass of 7.25$^{+1.19}_{-1.12}$~$M_{\oplus}$ based on archival and new radial velocity data. We also update the parameters of the longer-period, not-known-to-transit planet, and find it to be less eccentric and less massive than previously reported. The density of the transiting planet is $3.97$ g~cm$^{-3}$, which is low enough to suggest that the planet has at least a small volatile envelope, but the mass fractions of rock, iron, and water are not well-constrained. Given the host star brightness, planet period, and location of the planet near both the ``radius gap'' and the ``hot Neptune desert'', HD 86226c is an interesting candidate for transmission spectroscopy to further refine its composition. 

\end{abstract}

\section{Introduction}

The Transiting Exoplanet Survey Satellite (\tess{}) was launched in April 2018 and has proven to be a successful planet-finding machine, with more than 1100 new planet candidates having been detected so far.\footnote{NASA Exoplanet Archive, accessed 2020 February 20.} Compared to the Kepler mission, which ended in 2018, the \tess{} mission is observing much brighter stars, the brightest of which have already been targeted in radial-velocity (RV) surveys that stretch back a decade or more. The increased overlap between \tess{} and RV programs has allowed for faster validation of transiting planet candidates and immediate constraints on their masses \citep[e.g.,][]{Huang2018,Dragomir2019,Dumusque2019}. In turn, archival RV observations of \tess{} targets allow for more efficient vetting of the best planets for atmospheric spectroscopy. The initial interpretation of atmospheric observations requires at least $\pm$50\% mass precision, and more detailed analysis requires even higher-precision ($\pm$20\%) mass measurements \citep{batalha2019}. 

Such atmospheric observations are especially important for deducing the composition of planets in the ``super-Earth'' to ``sub-Neptune'' range, which are the most common type of planet with orbital periods shorter than $\sim$100 days \citep{howard2012,fressin2013,Fulton2017}. Mass and radius measurements alone suggest intrinsic astrophysical scatter in the 1-4~$R_{\oplus}$ planet mass-radius relation \citep[e.g.,][]{weiss&marcy2014,wolfgang2016}. Planets in this region of the mass-radius parameter space have degenerate distributions between core, mantle, water envelope, and/or H/He envelope -- especially for planets larger than $\sim$1.6~$R_{\oplus}$ \citep{rogers2015}, there is a range of combinations of iron, silicates, water, and gas that can match mass and radius measurements \citep{valencia2007b,adams2008,rogers&seager2010}. Interior structure models of small planets that incorporate constraints on the bulk refractory abundances of the host star have been shown to improve constraints on mantle composition, relative core size, ice mass fraction, etc., depending on the type of model and size of planet being modeled \citep{bond2010,carter-bond2012a,carter-bond2012b,dorn2015,Dorn2017,wang2019}, although counter-examples of different planet and host star abundance ratios certainly exist \citep[e.g.,][]{santerne2018}. Transmission and/or emission observations provide an additional window into the atmospheric compositions of small planets \citep{Miller-Ricci2009, Benneke2012,Morley2015, Diamond-Lowe2018} and even their surfaces \citep{Demory2016,koll2019,Kreidberg2019}.

Understanding the diversity in compositions of super-Earth and sub-Neptune exoplanets is important for tracing where and when they formed in protoplanetary disks, and for comparing their formation processes to those thought to have occurred in the solar system. For example, in our Solar System, Jupiter likely trapped solid material in the outer disk and created an inner mass deficit \citep[][]{Lambrechts2014,Kruijer2017}, thereby limiting the masses of the inner planets \citep[][]{Obrien2014,Batygin&Laughlin2015}. Jupiter's growth and possible migration are also thought to have scattered carbonaceous chondrite bodies inward, possibly delivering water and other volatile species to the Earth \citep[][]{Alexander2012,marty2012,OBrien2018}. Could analogous processes have happened in exoplanetary systems? How would this impact the compositions of inner super-Earth and sub-Neptune planets? From studies combining Kepler and long-term RV data, it appears that about one-third of small planets (1-4 $R_{\oplus}$ or 1-10 $M_{\oplus}$ inside 0.5 AU) have an outer giant planet companion (0.5-20 $M_{\rm{Jup}}$ and 1-20 AU), and that cool giant planets almost always have inner small planet companions \citep{zhu&wu2018,bryan2019}. To investigate whether outer giant exoplanets have a similar influence on inner small planets as they do in our Solar System, we can look for compositional differences between small inner planets that have outer giant planet companions versus those that do not.

In this paper, we describe the TESS detection of a small planet ($\sim$2~$R_{\oplus}$) around the solar-type star HD 86226, a system already known from RV studies to host a long-period giant planet. TOI-652.01 (HD 86226c) is a $\sim 4$days-period planet on the border between super-Earths and sub-Neptunes as defined by the ``radius gap'' identified by \cite{Fulton2017}. We also present a mass measurement of HD 86226c, and use all of the existing RV data to update the parameters of the known, $\sim 1600$day period giant planet. In \S2 we begin by reviewing what is known about the star and previously detected planet, and then in \S3 we detail the new or newly analyzed observations from TESS, Las Cumbres Observatory, All Sky Automated Survey (ASAS), ASAS for SuperNovae (ASAS-SN), Southern Astrophysical Research (SOAR) Telescope, CORALIE, and the Planet Finder Spectrograph (PFS). We present our analysis of the transit and RV observations in \S4, and our results in \S5. Finally, we discuss several interpretations of our results and summarize our conclusions in \S6. Overall, HD 86226c is an excellent candidate for atmospheric studies that can help address multiple small planet formation questions.

\section{Previous Characterization of HD 86226}

The proximity of HD 86226 and its similarity to the Sun have made it a target in many photometric, spectroscopic, and high-contrast imaging investigations, as well as planet searches. Here, we highlight the most relevant examples. Basic information about HD 86226 is listed in Table \ref{tab:stellarprop}. 

This star was included in the original Geneva-Copenhagen Survey (GCS) of the solar neighborhood \citep{nordstrom2004}, combining new RV measurements with existing $uvby\beta$ photometry, Hipparcos/Tycho-2 parallaxes, and proper motions to derive stellar parameters, kinematics, and Galactic orbits for a magnitude-limited sample of almost 16{,}700 F and G dwarfs. 
These derivations were subsequently improved with new Hipparcos parallaxes \citep{holmberg2009} to update the absolute magnitudes, ages, and orbits, and by using the infrared flux method \citep{casagrande2011} to update the stellar effective temperatures, metallicities, and ages. The final stellar parameters in the GCS reported by \cite{casagrande2011} for HD 86226 are as follows: $T_{\rm{eff}}$=5928$\pm$81 K, log~$g$=4.45 dex\footnote{No error provided}, [Fe/H] = 0.0 dex\footnote{No error provided}, $M_{*}$=1.05$^{+0.03}_{-0.05}$ $M_{\odot}$, and Age=3.31$^{+3.09}_{-2.3}$Gyr (using the BASTI isochrones, but the result is similar with Padova isochrones). Interestingly, \cite{datson2012,datson2015} conducted a focused study of solar analogues in the GCS, and confirmed that HD 86226 is not a solar twin, given the sufficiently great differences in its measured absorption line equivalent width (EW) values versus those measured from the solar spectrum.   
HD 86226 is included in the SWEET-Cat catalog of stellar parameters for stars with planets \citep{santos2013}, with the following parameters derived from high-resolution, high-signal-to-noise FEROS spectra based on the EWs of Fe I and Fe II lines, as well as the iron excitation and ionization equilibrium: $T_{\rm{eff}}$5947$\pm$21 K, log~$g$=4.54$\pm$0.04 dex, [Fe/H] = 0.02$\pm$0.02 dex. The most recently published stellar parameters of HD 86226come from \cite{maldonado2018} in their investigation of the compositional differences between stars hosting hot and cool gas giant planets. They reported  $T_{\rm{eff}}$=5854$\pm$13 K, log~$g$=4.36$\pm$0.03 dex, [Fe/H] = -0.05$\pm$0.01 dex,  as well as abundances of C, O, Na, Mg, Al, Si, S, Ca, Sci, Ti, V, Cr, and Mn. Along with their parameters, the Hipparcos $V$ magnitudes \citep{esa1997} and parallaxes \citep{vanleeuwen2007} and PARSEC isochrones \citep{bressan2012}, \citeauthor{maldonado2018} computed the mass, radius, and age of HD 86226 to be $M_{*}$=1.00$\pm$0.01 $M_{\odot}$, $R_{*}$=1.02$\pm$0.04 R$_{\odot}$, and Age=4.64$\pm$1.51 Gyr.

HD 86226 was also part of the long-term Magellan planet search program, initiated with the MIKE \citep{Bernstein2003} spectrograph on Magellan II fitted with an iodine absorption cell \citep{marcybutler1992,butler1996}. With thirteen RV observations from MIKE data spanning 6.5 yr, \cite{arriagada2010} reported the detection of HD 86226b, a 1534$\pm$280 days planet with an RV semi-amplitude $K$ of 37$\pm$15 m~s$^{-1}$ and a moderately high eccentricity of 0.73$\pm$0.21. The authors inferred the minimum mass of the planet to be $M_{rm{p}}$ sin~$i$=1.5$\pm$1.0 $M_{\rm{Jup}}$ (assuming $M_{*}$=1.02 $M_{\odot}$). However, an additional sixty-five RV observations from the CORALIE planet search \citep{udry2000} published in \cite{marmier2013} revealed that the long-period giant planet was less massive and less eccentric than previously thought. Those authors found $K=15.3\pm1.7$ m~s$^{-1}$, $e=0.15\pm0.09$, $P = 1695\pm58$ days, and $M_{\rm{p}}\sin i = 0.92\pm0.10 M_{\rm{Jup}}$ (assuming $M_{*}=1.06\pm0.03$ $M_{\odot}$).

There have been no published updates to the planet parameters of HD 86226b since \cite{marmier2013}. However, \cite{mugrauer&ginski2015} included the star in a search for  nearby stellar and substellar companions to known exoplanet host stars using the adaptive-optics imager NACO on the Very Large Telescope (VLT) at ESO's Paranal Observatory. With their $K_s$ band observations, the authors ruled out companions of $\geq$53 $M_{\rm{Jup}}$ between 12 and 387 AU around HD 86226. 

\begin{table}
  \caption{HD~86226 Catalog Information}
  \label{tab:stellarprop}
  \centering
  \small
  \begin{tabular}{p{0.2\linewidth}l c c}  
    \hline\noalign{\smallskip}
    HIP ID &  & 48739 \\
    TIC ID &  & 22221375 \\
    TOI ID &  & 652 \\
    R.A.$^{\mathrm{1}}$  & [J2000] & $09^h 56^m 29.84^s$  \\
    Dec.$^{\mathrm{1}}$ & [J2000] & $-24^{o}$ 05' 57.80'' \\
    $\mu_\alpha$$^{\mathrm{1}}$ & [mas/yr] & -177.11$\pm$0.09 \\
    $\mu_\delta$$^{\mathrm{1}}$ & [mas/yr] &  46.87$\pm$0.08 \\
    $\varpi^{\mathrm{1}}$ &  [mas]   &  21.86$\pm$0.050   \\
    RV$^{\mathrm{1}}$ & [km/s] & 19.56$\pm$0.19 \\
    $vsini^{\mathrm{5,6,7,8}}$ & [km/s] &  2.4-4 & \\
    SpT$^{\mathrm{2}}$    &  &  G2 V \\
    $V_T^{\mathrm{3}}$ & &8.004$\pm$0.013  \\
    $B_T$-$V_T^{\mathrm{3}}$    &  &  0.699$\pm$0.03 \\
    ${log(R^\prime_{HK})}^{\mathrm{4}}$ &  & -4.95 \\
    \noalign{\smallskip}\hline
  \end{tabular}
  \begin{list}{}{}
  \item[$^1$] \citet{Gaia:2018}; 
  $^2$ \citet{ESA1997_2} ; 
  $^3$ \cite{Hog:2000}; 
  $^4$ \citet{arriagada2011};
  $^5$ \cite{nordstrom2004};
  $^6$ \cite{marmier2013};
  $^7$ \cite{Glebocki&Gnacinski2005}
  $^8$ Also consistent with PFS measurement.
  \end{list}
\end{table}

\section{New observational characterization of HD 86226\label{sec:new_obs}}

    \subsection{Photometry}

    Here, we describe the details of the new photometric observations. Section \S4 contains the analysis of these observations.
        \subsubsection{TESS \label{sec:tess}}
        
        A new transiting planet candidate around HD 86226, TOI-652.01, was detected by the TESS Science Processing Operations Center (SPOC) pipeline \citep{Jenkins:2016} and announced in May 2019. The TESS short-cadence (2 minute) photometry was collected in Sector 9 (spanning 25.3 days from 2019 February 28 to 2019 March 26) using Camera 2. We downloaded the short-cadence lightcurve file from the Mikulski Archive for Space Telescopes (MAST), extracted the systematics-corrected photometry (PDCSAP\textunderscore FLUX; \citealt{Stumpe2012, Smith2012, Stumpe2014}), removed data points flagged as having low quality as well as out-of-transit 5$\sigma$ outliers, and normalized the lightcurve to have a mean value of unity outside of the transits (based on the period and transit duration reported in the SPOC Data Validation Report Summary for TOI-652.01).

\begin{figure*}
        \centering
        \includegraphics[scale=0.44]{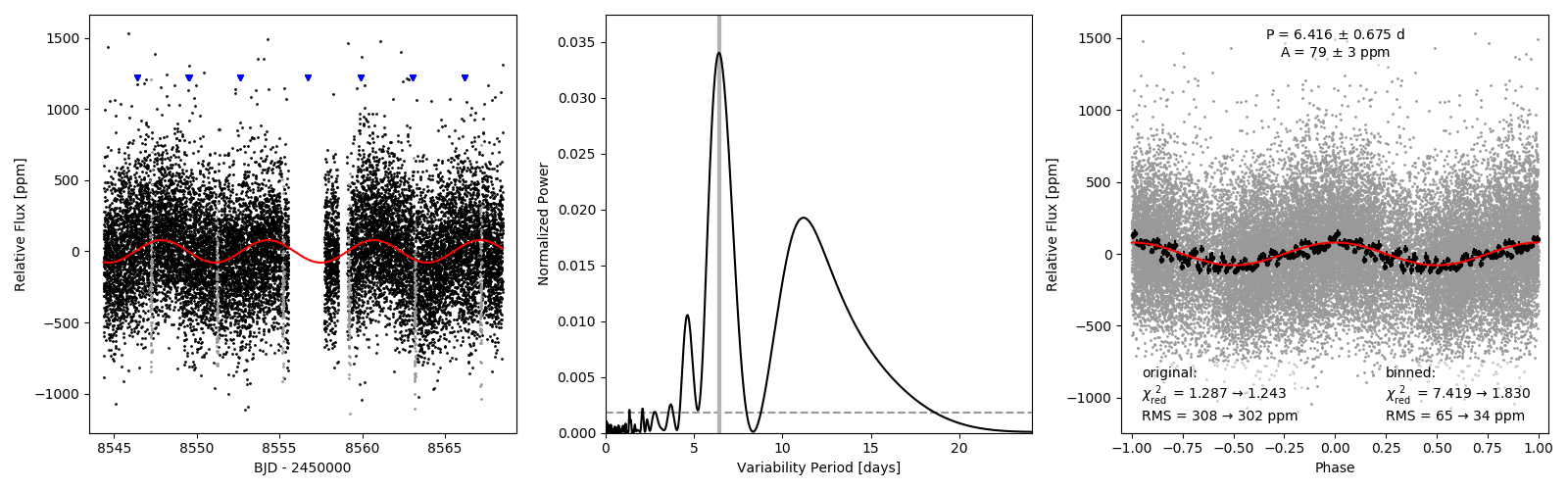}
        \caption{TESS photometric variability. Left panel: PDC lightcurve with transits masked (gray) and reaction wheel desaturation events marked with blue triangles. Middle panel: Lomb-Scargle periodogram with the strongest peak at 6.4 days marked with a vertical gray line. Horizontal dashed line represents a false alarm probability (FAP) of 0.01. Right panel: Phase-folded lightcurve at the variability period, with (black) and without binning (gray). Best-fitting variability function is shown in red.}
        \label{fig:tess_variability}
        \end{figure*}

        We detrended the lightcurve by modeling the low-order variability with a simple low-amplitude sine function, and then dividing by the sine function, which was derived by the following procedure. We masked the transits of TOI-652.01, then calculated a Lomb--Scargle (L--S) periodogram from the resulting lightcurve to determine the variability period and its error from the location and width of the highest L--S periodogram peak. The variability amplitude, error, and epoch were determined from the best-fitting sine function to the lightcurve, at the period determined from the L--S periodogram. The best-fit variability period is 6.4$\pm$0.7 days with an amplitude of 79$\pm$3 ppm, as shown in Figure \ref{fig:tess_variability}. We found that removing this variability reduced the binned photometry $\chi^2_{\rm red}$ from 7.42 to 1.83 and RMS from 65 to 34 ppm in the (L-S period) phase-folded lightcurve with 100 bins, comparable to the RMS of $\sim$31 ppm for the 10$^{th}$ percentile of the least noisy TESS lightcurves of similarly bright stars in Sector 9\footnote{TESS Data Release Notes for S9, \url{https://archive.stsci.edu/missions/tess/doc/tess\_drn/tess\_sector\_09\_drn11\_v04.pdf}}. We checked the 10 nearest stars in the short-cadence mode during Sector 9, ranging from between 484\arcsec and 3000\arcsec away from TOI-652, and detected no similar variability, although we cannot reject the possibility that the variability signal originates from a nearby diluted object. The period of the variability in TOI-652 is not near the reaction wheel desaturation events that happened every 3.125 days, and it is very close to perfectly sinusoidal, unlike the expected desaturation events. We conclude that the variability signal is likely of astrophysical origin, possibly half the rotation period or originating from another nearby diluted star. Our analysis of the ground-based photometry and activity indices below do not shed conclusive light on the signal's origin.

\begin{figure*}
        \centering
        \includegraphics[scale=0.8]{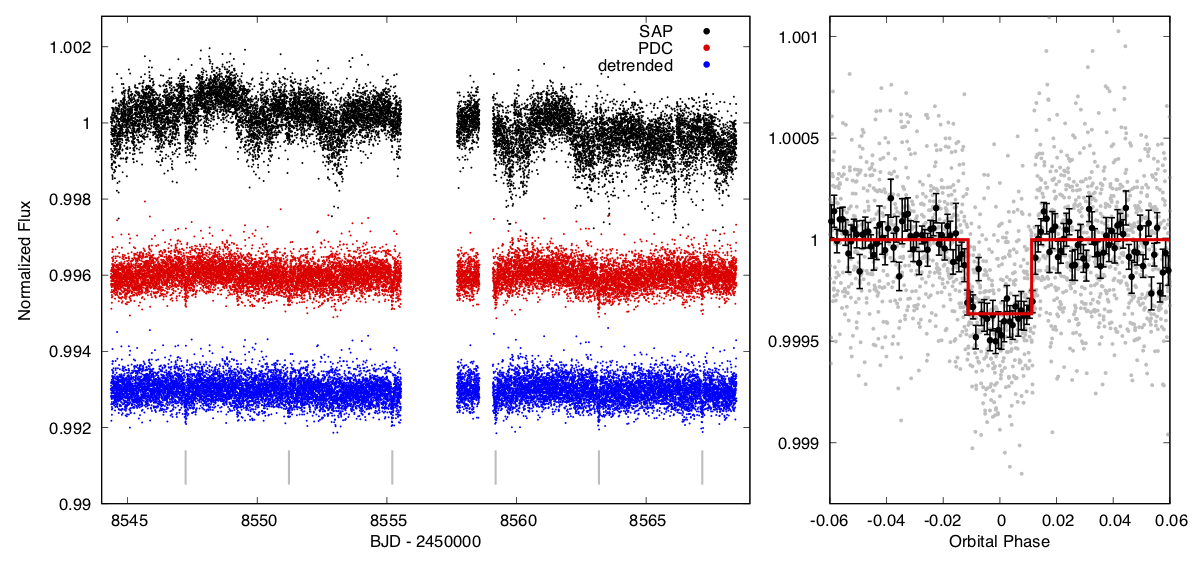}
        \caption{TESS lightcurve of HD 86226 and our preliminary transit detection. The left panel shows the simple-aperture (SAP, top, black), systematics-corrected (PDC, middle, red), and detrended (bottom, blue) {\it TESS} photometry, with the transits of TOI-652.01 (HD 86226c) marked with grey vertical lines at the bottom of the panel. The right panel shows the detrended lightcurve in gray, and in phase bins of 0.001 in black, folded to the period of TOI-652.01 as identified by the BLS periodogram. The detected transit is shown as a red line in the right panel.}
        \label{fig:tess_photometry}
        \end{figure*}

        We computed the box-fitting least squares (BLS) \citep{Kovacs:2002} periodogram of the detrended TESS lightcurve and independently detected a periodic transit signal at 3.98 days with a transit depth of 360 ppm and a signal-to-noise ratio of 20. This signal corresponds to TOI-652.01, reported by the SPOC pipeline. Next, we masked the detected transits and recomputed the BLS to search for any additional transiting planets in the system. The residual BLS periodogram peaks were all below 5$\sigma$ and none of the main residual peaks revealed a transit-like signal when phase folding the lightcurve at the corresponding periods; this is consistent with the Data Validation Report, which failed to find evidence of any additional transiting planet signatures in the lightcurve. In Figure \ref{fig:tess_photometry} we show the TESS photometry with transits of TOI-652.01 marked in the left panel, and our BLS-detected transit in the right panel.

        The Data Validation Report shows that TOI-652.01 passes all validation tests (such as the odd-even transit depth test for eclipsing binaries, the ghost diagnostics for scattered light, and the bootstrap false alarm test), except the centroid test of the difference image between the mean out-of-transit and in-transit images. The pixel response function centroid in the difference image relative to the TIC position and relative to the out of transit centroid are both offset by 12\arcsec (5$\sigma$) and in the same direction. However, the centroid offsets were likely caused by a slight saturation of the bright $T=7.4$ target, and because there are no known stars near the subpixel centroid displacement, it is reasonable to assume that the transit events happen on the target.

        The optimal photometric extraction mask was roughly two pixels in radius, and consisted of 17 pixels in total. Besides our target, seven other, much fainter stars were also located within the extraction mask and therefore contributed to the combined flux. The SPOC performs a correction for crowding and for the finite flux fraction of the target star's flux in the photometric aperture using the point-spread functions (PSFs) recovered during \tess{} commissioning; for TOI 652, this CROWDSAP value is 0.9985, indicating that the total contamination ratio based on the local background stars in the TIC is 0.15\%. Among the seven contaminating stars, only one star (TIC~22221380, 38.9\arcsec away) was bright enough ($T=14.88$) that it could be responsible for the observed 360 ppm transit depths in the combined flux if it were an eclipsing binary with eclipse depths of 36\%. However, the odd-even transit depth and the centroid tests did not indicate that TIC~22221380 was the likely source of the observed periodic brightness dips. The Las Cumbres observations described in \ref{sec:lcogt} also rule out an eclipse on TIC~22221380. Moreover, the RV follow-up observations (\S\ref{subsec:rvs}), which excluded TIC~22221380, revealed that the planetary signal comes from the main target, TIC~22221375. The small contamination from the nearby stars has been accounted for by the Presearch Data Conditioning (PDC) pipeline module \citep{Jenkins:2016}, which corrects for the dilution effect and also for any spilled flux of the target outside the photometric extraction mask.

        \subsubsection{Las Cumbres Observatory: Sinistro \label{sec:lcogt}}
        
        We acquired ground-based time-series follow-up photometry of a full transit of TOI-652.01 on UTC 2019 May 15 in $z$-short band from a Las Cumbres Observatory Global Telescope (LCOGT) 1.0 m telescope \citep{Brown:2013} at Siding Spring Observatory. We used the {\tt TESS Transit Finder}, which is a customized version of the {\tt Tapir} software package \citep{Jensen:2013}, to schedule the observations. The $4096\times4096$ LCO SINISTRO cameras have an image scale of 0$\farcs$389 pixel$^{-1}$ resulting in a $26\arcmin\times26\arcmin$ field of view. The 269 minute observation used 70 s exposure times, resulting in 162 images. The images were calibrated by the standard LCOGT BANZAI pipeline and the photometric data were extracted using the {\tt AstroImageJ} ({\tt AIJ}) software package \citep{Collins:2017}. Since the transit depth of TOI-652.01 is too shallow to generally detect with ground-based photometric follow-up observations, we saturated the target star in order to enable a search of the faint nearby Gaia DR2 stars for Nearby Eclipsing Binary (NEB) events that could have produced the \tess\ detection in the irregularly shaped \tess\ aperture that generally extends $\sim1\arcmin$ from the target star. A neighboring star that is fully blended in the \tess\ aperture and that is fainter than the target star by 8.6 magnitudes in \tess\ band and that has a 100\% eclipse could produce the \tess\ reported flux deficit at mid-transit. We therefore searched all stars within $2\farcm5$ that have a delta-magnitude $<9$ for deep eclipse events occurring within $3\sigma$ of the predicted time of transit center. All such nearby stars had lightcurves consistent with being flat with RMS values less (by at least a factor of eight) than the eclipse depth required to produce the \tess\ detection in each of the stars. By process of elimination, we conclude that the \tess\ detected transit is indeed occurring in TOI-652.01, or a star so close to TOI-652.01 that it was not detected by Gaia.

         \subsubsection{ASAS and ASAS-SN}

        We found archival $V$-band photometry of HD 86226 from the ASAS\footnote{\url{http://www.astrouw.edu.pl/asas/?page=aasc&catsrc=asas3}} \citep[ASAS;][]{Pojmanski1997} and from the ASAS-SN\citep{Shappee2014} database \footnote{\url{https://asas-sn.osu.edu}}. There are 660 ASAS measurements from 2000 November 21 to 2009 December 3. We selected the data flagged as ``A'' and ``B'', indicating the highest-quality measurements; data flagged as either ``C'' or ``D'' were automatically discarded. After performing a 3$\sigma$ clipping rejection, we ended up with 615 useful observations. ASAS-SN data spans from 2014 May 7 to 2018 July 5 with a total of 279 observations. In this case, we only performed the sigma-clipping process, yielding 274 measurements.

\begin{figure}
        \centering
        \includegraphics[width=\columnwidth]{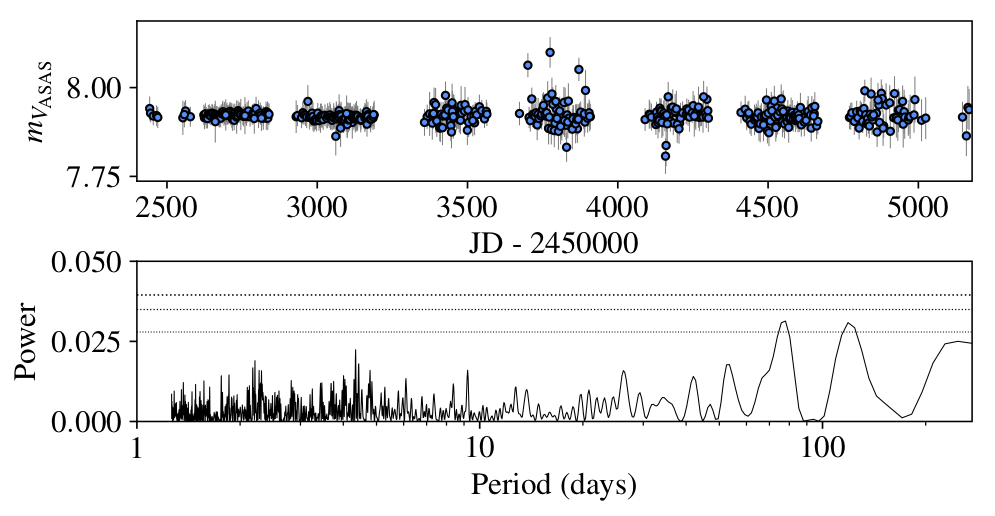}
        \caption{GLS periodogram of the ASAS $V$-band photometry. Dashed lines from top to bottom represent the 0.1,1 and 10\% significance threshold levels, respectively, obtained via 10,000 bootstraps samples.}
        \label{fig:asas_per}
        \end{figure}

\begin{figure}
        \centering
        \includegraphics[width=\columnwidth]{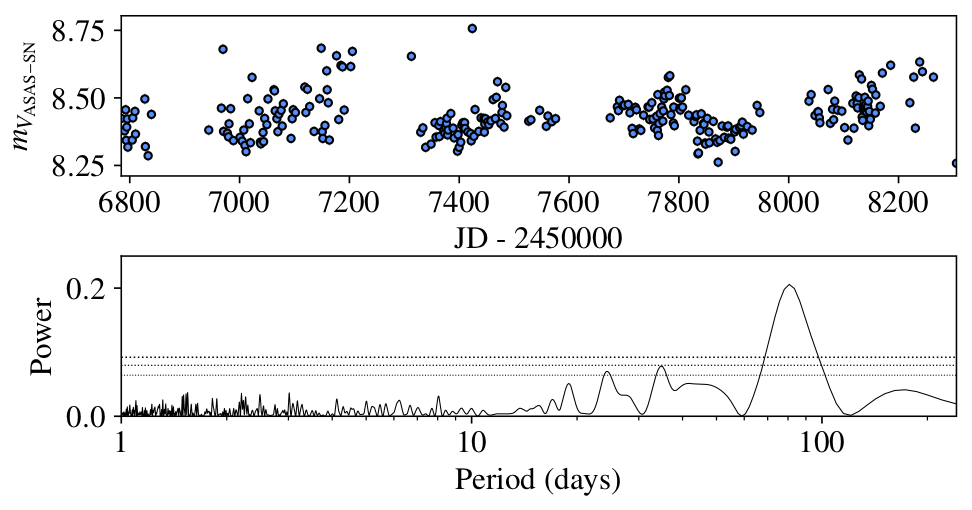}
        \caption{GLS periodogram of the ASAS-SN $V$-band photometry. Dashed lines from top to bottom represent the 0.1,1 and 10\% significance threshold levels, respectively, obtained via 10,000 bootstrap samples.}
        \label{fig:asassn_per}
        \end{figure}

        For both the ASAS and ASAS-SN photometry, we then applied the GLS periodogram to search for periodic signals embedded in the data that could be related to the rotation period of the star. We defined a grid of 20{,}000 period samples from 0.5 to 1000 days, evenly spaced in frequency space. The significance threshold levels were estimated in both cases by running 10,000 bootstraps on the input measurements via the \texttt{Python} module \texttt{astropy.stats.false\_alarm\_probability()}\footnote{\url{https://docs.astropy.org/en/stable/timeseries/lombscargle.html}}. Figures \ref{fig:asas_per} and \ref{fig:asassn_per} show the time series and GLS periodograms for the ASAS and ASAS-SN photometry, respectively. In both cases, the power spectrum shows a peak at $\sim$78 days, with the ASAS photometry also showing a peak slightly longward of 100 days. However, we caution that this should not necessarily be interpreted as the rotation period of HD 86226; given its solar-like parameters, we expect its rotation period to be closer to 20-30 days \citep{McQuillan2014}. \citet{arriagada2011} reported a rotation period of 25 days from analysis of chromospheric activity indicators (S-indices; see \S\ref{subsubsection:magellanpfs}) 
        of HD 86226, derived from their Magellan II/MIKE observations. In our analysis of the ASAS and ASAS-SN ground-based photometric data, we do not detect any significant peaks in the periodogram at/close to 25 days. Given the typical precision of the ASAS and ASAS-SN photometry (0.02-0.04 mag), we do not expect to detect the 79 ppm, 6.4 day \tess{} signal described in \S3.1.1.

        \subsubsection{WASP-South photometry}
\begin{figure}
        \centering
        \includegraphics[width=\columnwidth]{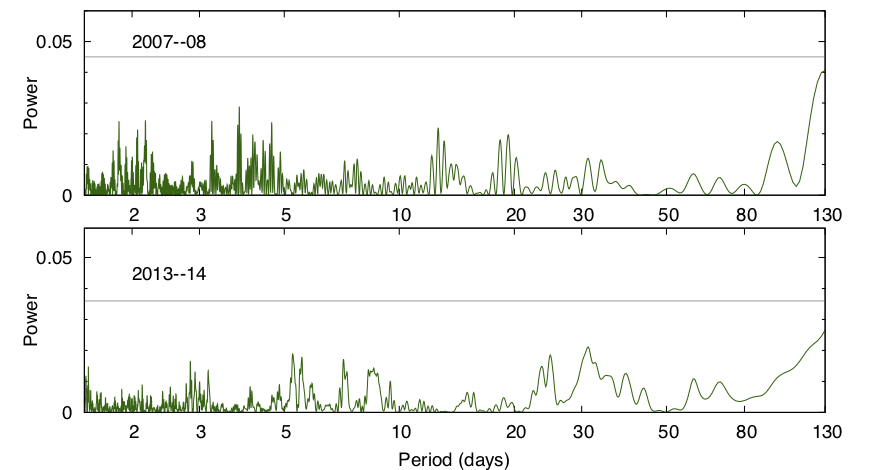}
        \caption{Periodogram of the WASP-South data from the 200 mm lenses (top) and 85 mm lenses (bottom). The horizontal lines are the estimated 1\%\ false-alarm levels.}
        \label{fig:wasp}
        \end{figure}

    WASP-South is the southern station of the WASP transit-search survey \citep{2006PASP..118.1407P}, and consisted of an array of eight cameras observing fields with a typical cadence of 10 min. The field of HD 86226 was observed over spans of 150 nights each year in 2007 and 2008, during which WASP-South was equipped with 200 mm, $f$/1.8 lenses, and then again in 2013 and 2014, equipped with 85 mm, $f$/1.2 lenses.  In all, 45,000 photometric observations were obtained. We searched the data for any rotational modulation using the methods from \citet{2011PASP..123..547M}.  We do not find any significant periodicity in the range 2-100 days, with a 95\% confidence upper limit on the amplitude of 2 mmag (Fig.~\ref{fig:wasp}).

    \subsection{High-resolution Imaging with SOAR}

\begin{figure}[h]
    \centering
    \includegraphics[scale=0.2]{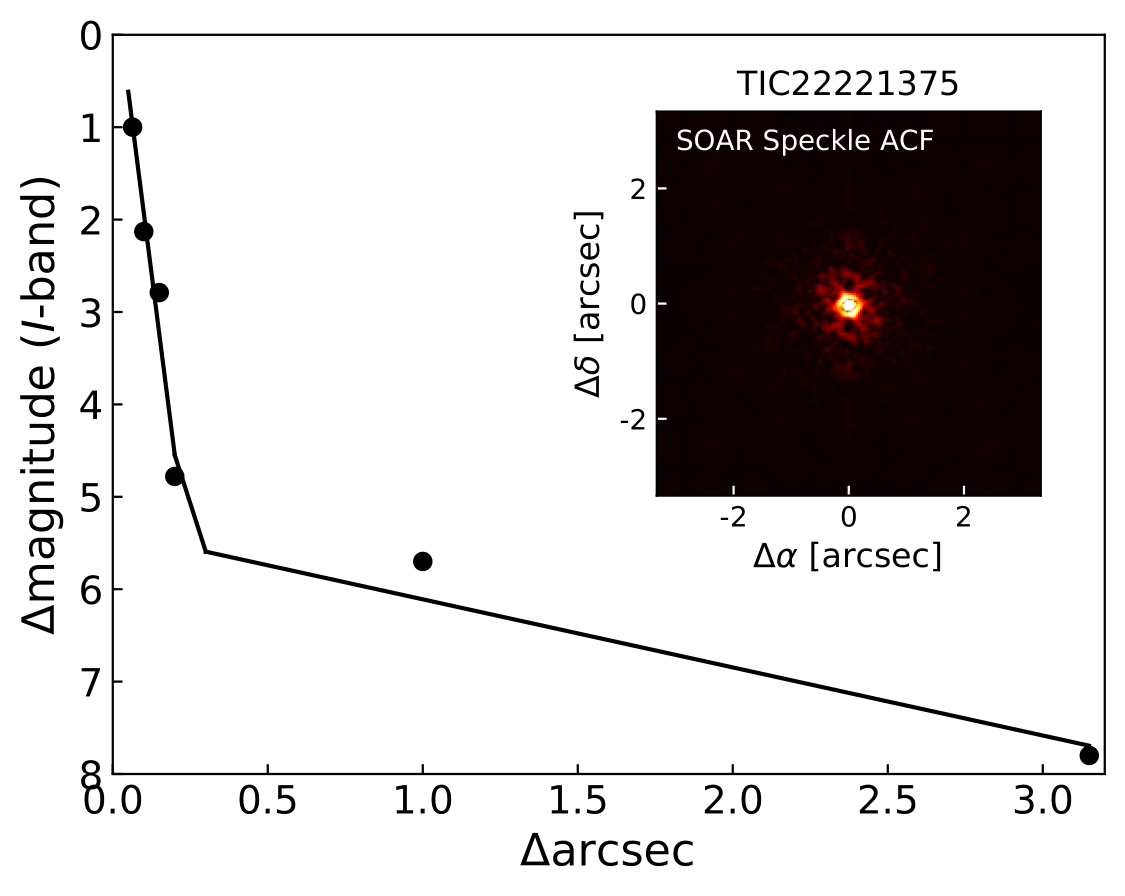}
    \caption{SOAR speckle results of HD 86226, showing the detection limit curve and reconstructed image (inset).}
    \label{fig:soar}
    \end{figure}

    With a very wide point-spread function ($\sim$1 arcmin), the TESS photometry may include flux from previously unknown nearby sources, including potential stellar companions, which can dilute the observed transit signal, resulting in an underestimated planetary radius. We searched for close companions to HD 86226 with speckle imaging on the 4.1 m SOAR telescope \citep{tokovinin2018} on 2019 May 18 UT. Observations were performed in the Cousins-I passband, similar to that of the TESS observations. No nearby stars were detected within 3\arcsec of the planetary host, and the data are able to place an upper limit of $\Delta$I$\sim6$ for any companions outside of 0.5\arcsec of the star. Within that separation, the data do not rule out brighter ($\Delta$I$<6$) unresolved companions. The 5$\sigma$ detection sensitivity and autocorrelation function of the latter observations are shown in Figure \ref{fig:soar}.

    \subsection{Spectroscopy}\label{subsec:rvs}
    Here, we describe the details of the new RV observations. Section \S4 contains the analysis of these observations to derive the mass of TOI-652.01 (hereafter HD 86226c).

        \subsubsection{Euler/CORALIE}
        HD 86226 has been monitored by the high-resolution spectrograph CORALIE \citep{CORALIE} on the Swiss 1.2 m Euler telescope in La Silla Observatory, Chile, starting in March 1999. Since the report by \cite{marmier2013}, 27 new CORALIE spectra have been acquired, making for a total of 78 RVs spanning 20 yr. The instrument underwent major upgrades in 2007 and 2014, which introduced offsets into the RV scale. For this reason, the CORALIE data presented in this study are treated as having come from three different instruments: CORALIE98, CORALIE07, and CORALIE14. The present version, CORALIE14, has a resolving power of $R\sim60,000$ and is fed by two fibers: one 2\arcsec\ diameter on-sky science fiber encompassing the star, and another that can either be connected to a Fabry-P\'{e}rot etalon for simultaneous wavelength calibration (used in the case of HD 86226) or on-sky for background subtraction of the sky flux. For bright stars such as HD 86226, CORALIE14 can reach its noise floor precision of 3 m~s${-1}$.

        We computed RVs for each epoch by cross-correlating the spectra with a binary G2 mask \citep{Pepe2002}. Bisector span, FWHM, and activity indicators were computed as well using the standard CORALIE DRS. All the new RVs presented here come from CORALIE14 and are presented in Table \ref{tab:COR}. The periodogram of H$\alpha$ activity index values \citep{Boisse+2009} derived from all of the existing CORALIE spectra is shown in black in the top of Figure \ref{fig:s_pfs}, with the window function in red.\footnote{Given the low signal-to-noise in the blue part of the CORALIE spectrum, we chose to use H$\alpha$ as our activity metric, instead of the S-index derived from Ca~{\sc ii} H and K lines.} We see one significant peak in the periodogram around 750 days, but it also overlaps partially with the window function.  

        \begin{table}
	   \centering
	    \caption{New CORALIE Radial Velocities for HD 86226}  
	    \begin{tabular}{lccc} 
	    BJD 	&	RV & $\sigma$RV & H$\alpha$-index\\
	    \hline
	    (- 2450000) & (-19700 \mps) & (\mps)&  \\
	    \hline 
        7024.75050 & 60.88 & 3.02 & 0.187 \\
        7038.86008 & 59.01 & 4.06 & 0.190  \\
        7118.65893 & 59.31 & 6.58 & 0.187 \\
        7143.58148 & 67.54 & 2.99 & 0.187 \\
        7430.82012 & 49.40 & 10.94 & 0.188 \\
        7459.76833 & 58.61 & 4.43 & 0.189 \\
        7739.79877 & 54.72 & 3.64 & 0.189 \\
        ...&...&...&...\\
	    \hline
	    \end{tabular}
	    \tablecomments{This table is published in its entirety in the machine-readable format. A portion is shown here for guidance regarding its form and content.}
        \label{tab:COR}
        \end{table}

        \subsubsection{Magellan II/PFS}\label{subsubsection:magellanpfs}
        HD 86226 has been monitored as part of the Magellan Exoplanet Search, first with MIKE \citep{Bernstein2003} as described above, and more recently with the Planet Finder Spectrograph \citep[PFS;][]{crane2006,crane2008,crane2010} as one of the targets of the long-term survey. Each observation makes use of the iodine cell. A template spectrum of the target star (without the iodine cell and at $R\sim 130,000$) is needed for the computation of the RVs. We bracket these template observations with the spectra of a rapidly rotating B star that is used in the determination of the instrumental profile (PSF), which is also necessary for the forward-modeling process. The determination of precise RVs follows an updated version of the steps described by \cite{butler1996}. The PFS detector was upgraded in February 2018. To take into account the change in the velocity zero-point offset between the different setups, throughout the analysis we refer to the data prior to the CCD upgrade as PFS1 and the post-fix data with the new detector as PFS2. Much of the PFS1 iodine data were observed with a 0.5\arcsec slit, resulting in a resolving power of $\sim$80,000, whereas the PFS2 data were observed with a 0.3\arcsec slit, resulting in a resolving power of $\sim$130,000. Exposure times for HD 86226 ranged from roughly 5-15 minutes with PFS1 and 10-20 minutes with PFS2.
        
\begin{figure}
            \centering
            \includegraphics[width=\columnwidth]{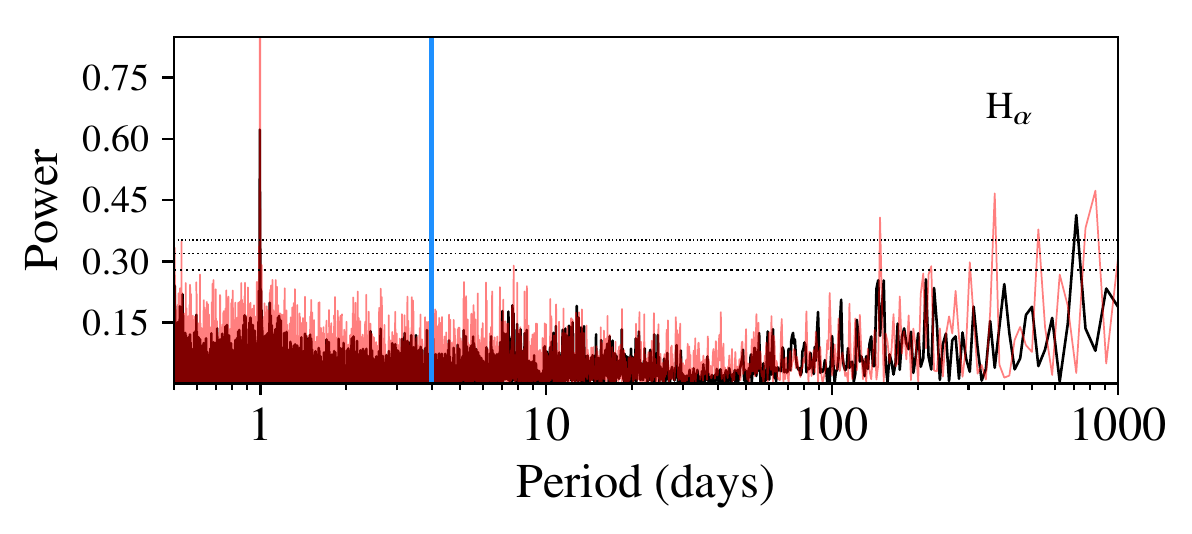}
            \includegraphics[width=\columnwidth]{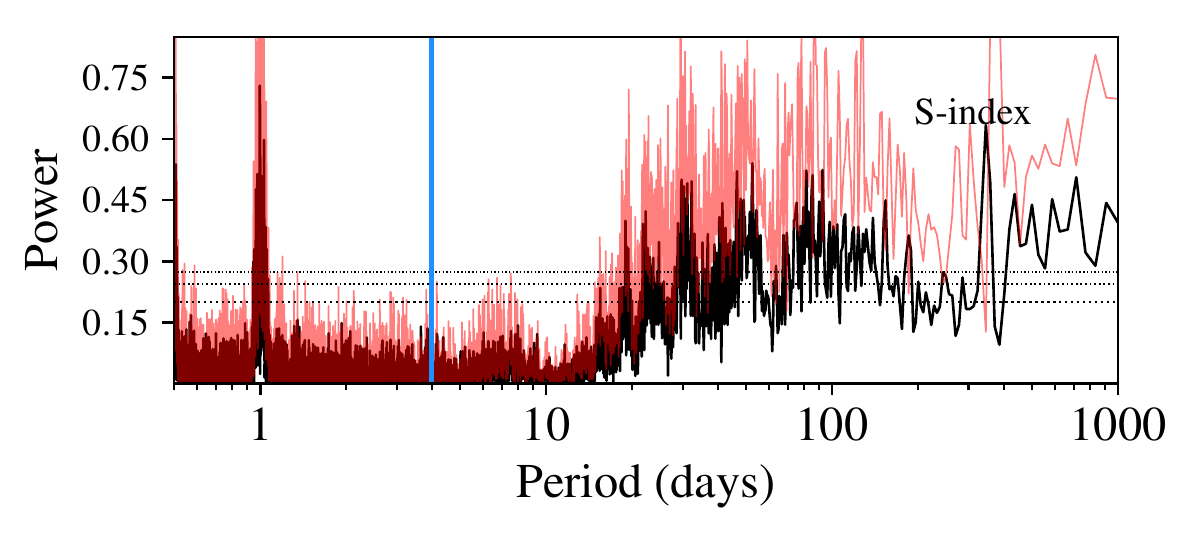}
            \caption{Top: Generalized Lomb-Scargle (GLS) periodogram for H$_{\alpha}$ indices from CORALIE. Bottom: GLS of S-indices from PFS. In both panels, the blue vertical line represents the position of the 3.9-day period of the planet candidate. The red power spectra show the window function.}
            \label{fig:s_pfs}
        \end{figure}

        In an attempt to monitor the chromospheric activity of the star, we also derive spectroscopic indices from the Ca~{\sc ii} H line (S-index; after \citealt{Wright2004}) 
        with our reduction pipeline. Figure \ref{fig:s_pfs}, bottom, shows the periodogram of the S-indices derived from the PFS spectra, in which we see a collection of significant peaks starting at $\sim$50 days and continuing to longer periods, with a noticeable peak at $\sim$345 days. Again, many of these peaks also overlap with the window function, shown in red. 
        
        Table \ref{tab:pfsrvs} shows the RV measurements for HD 86226 acquired with PFS. The estimated parameters of HD 86226c met the criteria for inclusion in the target list of the Magellan-TESS Survey (MTS), a TESS follow-up program to measure the masses of $\sim$30 R$_{\rm{p}} \leq 3$~$R_{\oplus}$ planets to construct an unbiased mass-radius relation and investigate the relationship between small planet density and insolation flux, host star composition, and system architecture. More details of the survey will be published in a forthcoming paper (Teske et al. 2020, in preparation). We include in this publication a total of 105 individual spectra spanning $\sim$9.4 yr, from 2010 January 2 to 2019 May 24; as a result of HD 86226c being included in MTS, our observing cadence increased during the 2019 May observations.   
        
        \begin{table}
	        \centering
	        \caption{PFS Radial Velocities for HD 86226}
	        \begin{tabular}{lcccc} 
	        BJD$^{a}$ 	&	RV & $\sigma_{\rm RV}$ & S-index &Note\\
	        \hline
	        (- 2450000)& (\mps) & (\mps)&&\\
	        \hline 
            5198.80046& ~4.20& 1.64& 0.165&PFS1\\
            5198.80272& ~9.66& 1.79& 0.171&PFS1\\
            5252.70830& ~5.98& 1.59& 0.157&PFS1\\
            5256.67966& ~3.18& 1.45& 0.193&PFS1\\
            5339.53760& 12.55& 1.47& 0.159&PFS1\\
            5339.54366& 14.21& 1.29& 0.161&PFS1\\
            5581.81336& ~3.98& 1.42& 0.165&PFS1\\	
            ...&...&...&...&...\\
	        \hline
	        \end{tabular}
	        \tablenotetext{a}{These dates were converted from JD$_{\rm UTC}$ to BJD$_{\rm TDB}$ using PEXO \citep{Feng2019}.}
	    \tablecomments{This table is published in its entirety in the machine-readable format. A portion is shown here for guidance regarding its form and content.}
        \label{tab:pfsrvs}
        \end{table}

\section{Analysis}
    
    \subsection{Stellar Characterization}
        
\begin{figure}[h]
    \centering
    \includegraphics[width=0.43\textwidth,trim={0cm 8 0 0},clip]{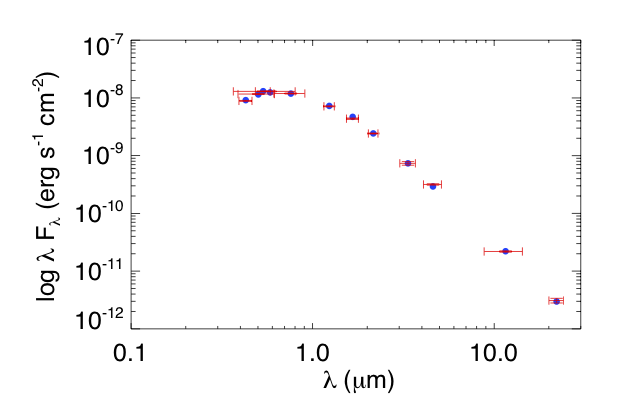}
    \includegraphics[width=0.4\textwidth,trim={0cm 5 0 5},clip]{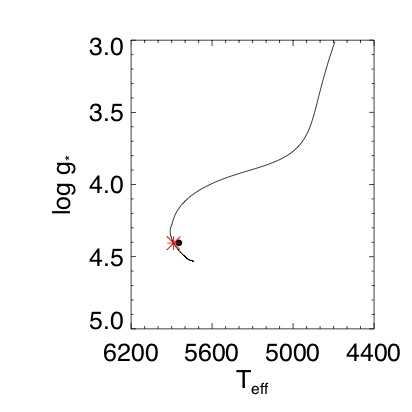}
    \caption{SED (top, see Table \ref{tab:sed}) and fit to MIST stellar evolutionary model (bottom) for HD~86226. The resulting parameters are listed in Table \ref{tab:exofast}.}
    \label{fig:sed_isochrone}
    \end{figure}

    We used EXOFASTv2 \citep{Eastman:2013, Eastman:2017} to fit a spectral energy distribution (SED) and the MIST stellar evolutionary models \citep{Dotter:2016} to determine the stellar parameters, as shown in Figure \ref{fig:sed_isochrone}. We used spectroscopic priors on the effective temperature and [Fe/H] of 5854$\pm$50 K and -0.05$\pm$0.08 dex, respectively, from \citet{maldonado2018}, rounding their quoted uncertainties up to account for systematic error floors. We also applied a parallax prior of 21.943$\pm$0.060 mas from Gaia DR2 \citep{Gaia:2018}, after applying the 0.082 mas systematic offset determined by \citet{Stassun:2018} and an upper limit on the $V$-band extinction of 0.15097 mag from \citet{Schlafly:2011}. The SED fit was performed using an SED fitting code \citep{Eastman2019} that interpolates the 4D grid of log~$g$, $T_{\rm eff}$, [Fe/H], and extinction grid from Conroy et al. 2020, (in preparation)\footnote{\url{http://waps.cfa.harvard.edu/MIST/model_grids.html\#bolometric}} to determine the bolometric corrections in each of the observed bands, summarized in Table \ref{tab:sed}. This version of the code is more accurate than the currently public version for wide bandpasses like Gaia because it accounts for the detailed shape of the filter. Note that the SED fitting takes into account errors in the zero points of the filters, but not systematics in the stellar atmospheric models and their calibration to stellar data. Thus, we round up the errors on $R_{*}$ to 2.5\% and and $T_{\rm eff}$ to 1.5\%. Our resulting parameters in Table \ref{tab:exofast} are consistent (equivalent within errors) with previous derivations for this star using both photometric and spectroscopic data \citep[e.g.,][]{holmberg2009,casagrande2011,marmier2013,santos2013,Gaia:2018}, although we note that our parameters place HD 86226closer to spectral type G1~V versus G2~V \citep{pecaut&mamajek2013}.

    \begin{table}[t]
        \caption{HD~86226 Photometry}
        \label{tab:sed}
        \centering
        \footnotesize
        \begin{tabular}{l c c c c}  
        \hline\noalign{\smallskip}
        Band & Mag & Used   & Catalog  & Catalog \\
            &     &    mag error       &    mag error    & \\
        \hline\noalign{\smallskip}
        $B_T$& 8.703 &0.020 &0.019 & \cite{Hog:2000}\\
        $V_T$ & 8.004 &0.020 &0.013 &\cite{Hog:2000}\\
        $J_{2M}$ & 6.839 & 0.020 &  0.020 & \citet{Cutri:2003}\\
        $H_{2M}$ & 6.577 &0.030& 0.030 & \citet{Cutri:2003}\\
        $K_{2M}$ & 6.463 &0.020 &0.020& \citet{Cutri:2003}\\
        WISE1 & 6.446 &0.078& 0.078& \citet{Cutri:2013}\\
        WISE2 & 6.377& 0.030 &0.024& \citet{Cutri:2013}\\
        WISE3&6.447 &0.030 &0.016& \citet{Cutri:2013}\\
        WISE4&6.392 &0.100 &0.071& \citet{Cutri:2013}\\
        Gaia &7.771 &0.020 &0.001& \citet{Gaia:2018}\\
        Gaia$_{\rm{BP}}$ &8.108 &0.020& 0.002& \citet{Gaia:2018}\\
        Gaia$_{\rm{RP}}$ &7.333 &0.020 &0.004& \citet{Gaia:2018}\\
        \noalign{\smallskip}\hline
    \end{tabular}
    \end{table}
        
    \begin{deluxetable}{lcc}
    \tablecaption{Median Values and 68\% Confidence Interval for HD~86226 \label{tab:exofast}}
    \tablehead{\colhead{~~~Parameter} & \colhead{Units} & \multicolumn{1}{c}{Values}}
    \startdata
    \multicolumn{2}{l}{Stellar Parameters:}&\smallskip\\
    ~~~~$M_*$ &Mass (\msun) &$1.019^{+0.061}_{-0.066}$\\
    ~~~~$R_*$ &Radius (\rsun) &$1.053^{+0.026}_{-0.026}$\\
    ~~~~$L_*$ &Luminosity (\lsun) &$1.180^{+0.036}_{-0.029}$\\
    ~~~~$\rho_*$ &Density (g~cm$^{-3}$) &$1.226^{+0.092}_{-0.091}$\\
    ~~~~$\log{g}$ &Surface gravity (dex, with $g$ in cm~s$^{-2}$) &$4.400^{+0.029}_{-0.032}$\\
    ~~~~$T_{\rm eff}$ &Effective Temperature (K) &$5863^{+88}_{-88}$\\
    ~~~~$[{\rm Fe/H}]$ &Metallicity (dex) &$0.018^{+0.057}_{-0.043}$\\
    ~~~~$[{\rm Fe/H}]_{0}$ & Initial metallicity\tablenotemark{a}  &$0.040^{+0.055}_{-0.050}$\\
    ~~~~Age &Age (Gyr) &$4.6^{+3.7}_{-2.7}$\\
    ~~~~EEP &Equal evolutionary point\tablenotemark{b}  &$361^{+38}_{-28}$\\
    ~~~~$A_V$ & $V$-band extinction (mag) &$0.05^{+0.04}_{-0.03}$\\
    ~~~~$\sigma_{\rm{SED}}$ &SED photometry error scaling\tablenotemark{c}  &$1.72^{+0.55}_{-0.35}$\\
    ~~~~$\varpi$ &Parallax (mas) &$21.94^{+0.06}_{-0.06}$\\
    ~~~~$d$ &Distance (pc) &$45.57\pm0.12$\\
    \enddata
    \tablecomments{Created using EXOFASTv2 commit number 86bb5c9.}
    \tablenotetext{a}{This is the birth metallicity of the star, which is the input to the theoretical model. The measured surface metallicity evolves throughout the life of the star. For more details, see \cite{choi2016}.}
    \tablenotetext{b}{This represents a uniform basis to describe the evolution of all stars, such that each phase of stellar evolution is represented by a fixed number of points. For more details, see \cite{Dotter:2016} and \cite{choi2016}.}
    \tablenotetext{c}{Errors in the broadband photometry (Table \ref{tab:sed}) are scaled by this factor, which essentially enforces the model has a $\chi^2$/dof=1.}
    \end{deluxetable}


    \subsection{Transit Modeling}\label{subsec:transit_only}

    First, we wanted to further constrain the orbital period and center-of-transit time of HD 86226c. Using the exoplanet photometry and RV analysis code \texttt{juliet} \citep{Espinoza2018b}, we modeled the transit lightcurve from \S\ref{sec:tess} with the priors defined in Table \ref{tab:photometry_only_priors}, based on our initial BLS periodogram and the reported TESS SPOC pipeline results. We tried incorporating an additional correlated noise term in the form of a squared-exponential (SE) Gaussian process (GP) kernel, but found the models to be indistinguishable based on the model evidences ($\Delta\ln Z < 1$). This makes sense, given the low level of activity of the star. 
    Moving forward, we only account for white noise in the TESS lightcurve modeling for HD 86226c, although we keep the additional ``jitter'' term ($\sigma_{\rm{TESS}}$) added in quadrature to the reported photometric uncertainties to represent any residual signal. 

    \begin{table*}
        \centering
        \footnotesize
        \caption{Priors Used in TESS Photometry-only Fits with \texttt{juliet}. }  \label{tab:photometry_only_priors}
        \begin{tabular}{lcc}
            \hline
            \hline
            \noalign{\smallskip}
            Parameter & Prior & Description \\
            \noalign{\smallskip}
            \hline
        \textbf{no GP}& & \\
        Planet orbit parameters & & \\

        $P$    & $\mathcal{N}(3.9852,0.01)$ & Period of HD 86226c(d) \\
        $t0$  & $\mathcal{N}(2548543.2458,0.01)$ & Center-of-transit time for HD 86226c (d)\\
        $a$ &$\mathcal{N}(12.5,48)$ & Scaled semi-major axis ($a/R_{*}$) of orbit for HD 86226c \\
        $r1$ &$\mathcal{U}(0,1)$ & Parameterization of $p$ and $b$ for HD 86226c$^{a}$\\
        $r2$ & $\mathcal{U}(0,1)$ & Parameterization of $p$ and $b$ for HD 86226c$^{a}$\\
        $e$ & 0 (fixed) & Eccentricity of orbit for HD 86226c$^{b}$ \\
        $\omega$ & 90 (fixed) & Argument of periastron passage of HD 86226c$^{b}$ orbit (deg) \\
        TESS photometry parameters & & \\
        $q_1$ & $\mathcal{U}(0,1)$ & Quadratic limb-darkening parameterization$^{c}$ \\
        $q_2$ & $\mathcal{U}(0,1)$ & Quadratic limb-darkening parameterization$^{c}$ \\
        $D_{\rm{TESS}}$ &  1 (fixed) & Dilution factor for TESS photometry \\
        $M_{\rm{TESS}}$ & $\mathcal{N}(0.01,1000)$&  Relative out-of-transit target flux for TESS photometry\\
        $\sigma_{\rm{TESS}}$ & $\mathcal{J}(0.01,1000)$ & Offset relative flux for TESS photometry (ppm)\\

        \textbf{additional priors for squared-exponential GP kernel}  & & \\
        GP$_{\sigma}$ & $\mathcal{J}(1,100)$ & Amplitude of the GP (ppm) \\
        GP$_{\alpha_0}$ & $\mathcal{J}(1,100)$ & Inverse (squared) length-scale/normalized amplitude \\
        \hline
        \end{tabular}
    \begin{flushleft}
        \tablecomments{The labels ~$\mathcal{U}$, $\mathcal{N}$, and $\mathcal{J}$ represent uniform, normal, and Jeffrey's (log-normal) distributions, respectively.}
        \tablenotetext{a}{Here, $p$ is the planet-to-star radius ratio and $b$ is the impact parameter of the orbit. The \texttt{juliet} parameterization was proposed by \cite{Espinoza2018a} and only allows for physically possible values (i.e., $b < 1 + p$); see reference for details. We also defined the minimum and maximum planet-to-star radius ratios to be 0 and 1, respectively.}
        \tablenotetext{b}{We also tried allowing these parameters to float, only requiring that $e < 1$, and found no significant difference in the planet orbital period or center-of-transit time.}
        \tablenotetext{c}{From \cite{kipping2013}, the transformations from ($q_1$,$q_2$) to the quadratic limb-darkening coefficients ($u_1$,$u_2$) are $q_1=(u_1 + u_2)^{2}$ and $q_2=0.5~u_1 \times (u_1 + u_2)^{-1}$.}
    \end{flushleft}
    \end{table*}

    \subsection{RV Detection} \label{subsec:rvonly}

    \begin{figure}[h]
        \centering
        \includegraphics[width=\columnwidth]{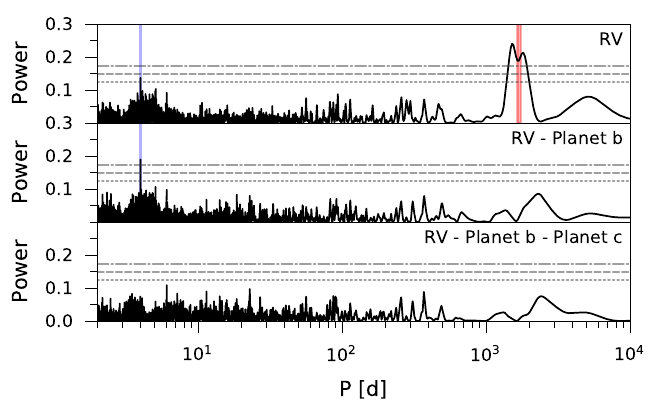}
        \caption{GLS periodogram of the RV dataset presented in \S\ref{subsec:rvs} of HD 86226. The red vertical line indicates the period of the known planet detected from RVs ($P = 1700\pm60$ days as measured by \citealt{marmier2013}) while the blue vertical line indicates the period of the transiting candidate reported by the TESS SPOC pipeline ($P = 3.98$ days). Top panel: GLS of the complete RV dataset after fitting from the individual RV offsets. Middle panel: GLS of the RV residuals after fitting a sinusoid to the highest peak in the top panel corresponding to the known Jupiter-like planet. Bottom panel: GLS of the RV residuals after fitting with a sinusoid the highest peak in the middle panel, corresponding to the transiting planet alerted by TESS. Horizontal lines show the FAP levels of 10\% (short-dashed line), 1\% (long-dashed line), and 0.1\% (dotted-dashed line) for each panel.}
     \label{fig:gls_rv}
    \end{figure}

    Next, we reanalyzed the available RV observations accounting for the inner transiting planet in addition to the already known Jupiter-like companion, and searched for possible additional planet signals. Figure~\ref{fig:gls_rv} shows the Generalized Lomb-Scargle \citep[GLS;][]{Lomb1976,Scargle1982,Zechmeister2009}  of the complete RV dataset presented in Sect.~\ref{subsec:rvs}. For each panel, we compute the theoretical false alarm probability (FAP) as described in \citet{Zechmeister2009}, and show the 10\%, 1\%, and 0.1\% levels. As seen in the top panel, the highest peak in the GLS corresponds to the known long-period planet, although the peak is wide and asymmetric meaning that its period is not well-constrained. However, the second highest peak in the periodogram is at $P = 3.98$ days, with a $\mathrm{FAP}<10\%$, corresponding to the reported transiting candidate from TESS data. Removing the RV signature of the known long-period planet (fit with a Keplerian orbit) from the RVs, we find that the peak at $P = 3.98$ days increased in significance and surpassed the $\mathrm{FAP}=0.1\%$ threshold, meaning that the planet might have been detected using the RV data even without any knowledge of the transit signal. After accounting for the signals produced by the two planets, no significant peaks remain in the periodogram.
 
    We used \texttt{juliet} to perform a systematic model comparison analysis and computed the Bayesian model log evidence ($\ln Z$) using the \texttt{dynesty} package \citep{dynesty}. As in \citet{Luque2019A&A...628A..39L}, we consider a model to be moderately favored over another if the difference in its $\ln Z$ is greater than two, and strongly favored if it is greater than five. If models are indistinguishable ($\Delta \ln Z < 2$), the one with fewer degrees of freedom is preferred. 

    Table~\ref{tab:models} shows the different models tested to fit the RV data together with the orbital period priors and the corresponding $\ln Z$ values. The rest of the priors of the fits are uniform and uninformative; the eccentricity and argument of periastron are sampled through the parameterization $\sqrt{e} sin(\omega)$, $\sqrt{e}cos(\omega)$ with boundaries between -1 and 1. First, we note that including the transiting planet in the fit improves the evidence of the model significantly, again suggesting that the transiting planet could have been detected using the RV data alone. To account for additional systematics or correlated noise in the data we include in the model an SE GP kernel\footnote{Squared-exponential (SE) GP kernel of the form $k_{i,j} = \sigma^2_\mathrm{GP,RV} \exp\left(-\alpha_0 (t_i - t_j)^2\right)$}. Since including a GP leads to an insignificant improvement in the model, we conclude that the simpler, two-planet model is the one that best describes the RV data.

    \subsection{Joint Transit$+$RV Fitting}\label{subsection:jointfit}

    We jointly analyzed the transit photometry and RV time series using the \texttt{juliet} package, which as mentioned above allows the user to fit the photometric time series and the RVs from multiple instruments at the same time. We set up a two-planet model consisting of a nontransiting planet and a transiting planet with parameters from the TESS alert. The priors for the orbital parameters for this joint fit were chosen according to the information from the previous transit-only and RV-only analyses (see Sections \ref{subsec:transit_only} and \ref{subsec:rvonly}), except that we allowed the eccentricity of HD 86226c to be a free parameter, using a normal distribution between 0 and 1 truncated at 0.2 (thus, $e>0.2$ is allowed at much lower probability). We performed the joint fit using the Dynamic Nested Sampling algorithm via the \texttt{Python} module \texttt{dynesty} \citep{Speagle2018,Speagle2019}.

\begin{table}
    \centering
    \caption{Model Comparison of RV-only Fits with \texttt{juliet}. }  \label{tab:models}
    \begin{tabular}{llcc}
        \hline
        \hline
        \noalign{\smallskip}
        Model & Prior $P_{\rm planet}$ (d) & GP kernel  & $\Delta\ln Z$    \\
        \noalign{\smallskip}
        \hline
        \noalign{\smallskip}
0pl     & \dots                                         & \dots     & 45.8      \\[0.1cm]
1pl     & $\mathcal{U}_\mathrm{b}(1350,1750)$           & \dots     & 15.0      \\[0.1cm]
{\bf 2pl}     & $\mathcal{U}_\mathrm{b}(1350,1750)$     & \dots     & {\bf ~0.6}      \\
        & $\mathcal{N}_\mathrm{c}(3.984,0.001^2)$       &           &          \\[0.1cm]
        \noalign{\smallskip}
        \noalign{\smallskip}
1pl+GP  & $\mathcal{U}_\mathrm{b}(1350,1750)$           & SE       & 14.2      \\[0.1cm]
2pl+GP  & $\mathcal{U}_\mathrm{b}(1350,1750)$           & SE       & ~0.0      \\
        & $\mathcal{N}_\mathrm{c}(3.984,0.001^2)$       &           &          \\[0.1cm]
        \noalign{\smallskip}
        \hline
    \end{tabular}
    \tablecomments{The prior labels $\mathcal{U}$ and $\mathcal{N}$ represents an uniform and normal distribution, respectively. The final model used for the joint fit is marked in boldface (see Sect.~\ref{subsec:rvonly} for details about the selection of the final model).}
 \label{tab:rv_only_posteriors}
\end{table}

\section{Results}

\begin{figure*}
    \centering
    \includegraphics[scale=0.85]{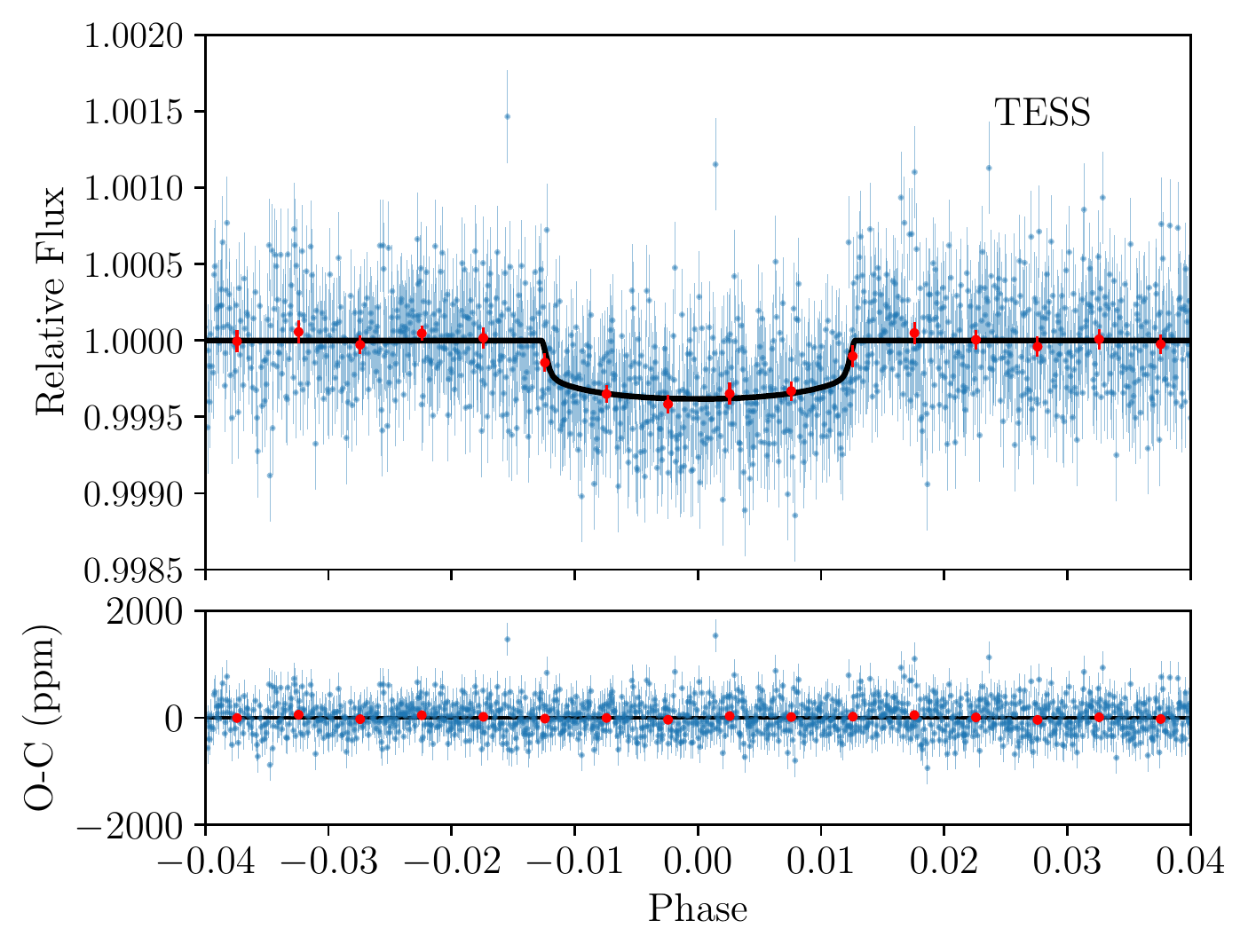}
    \includegraphics[width=\textwidth]{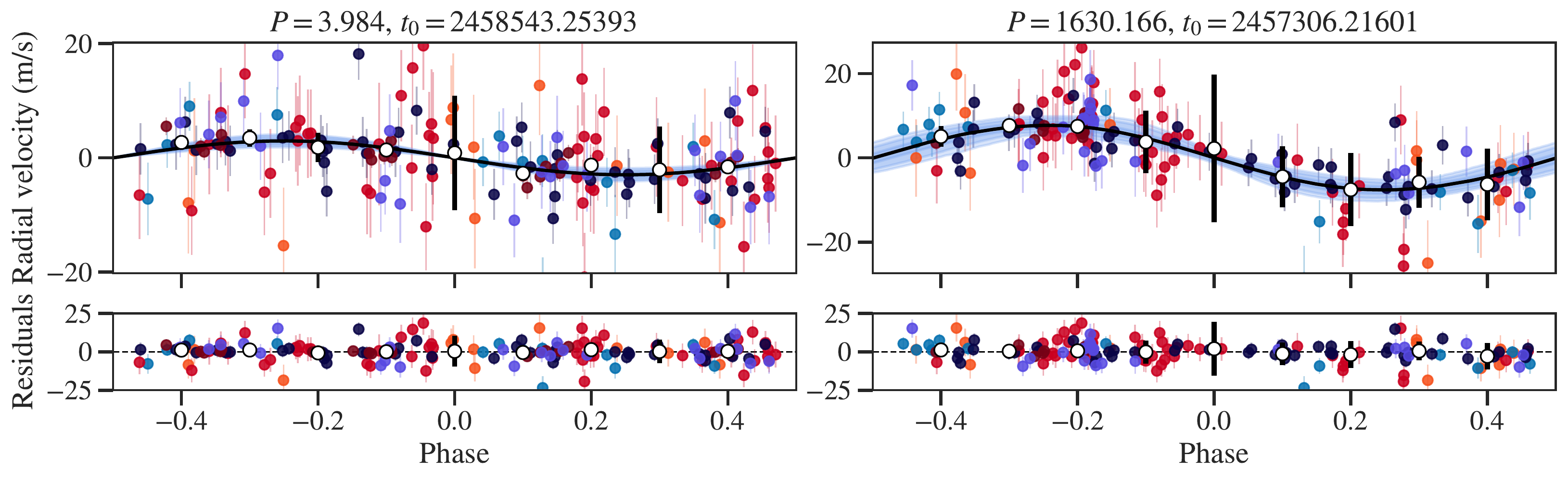}
    \caption{Results from our joint fit for a two-planet model using \texttt{juliet}. Top: Phased-folded lightcurve from {\it TESS} photometry (blue points) and transit model (black solid line) for the inner planet. Red points show binned photometry using a bin of 0.005 in phase. Residuals are shown in the bottom panel. Bottom: Phase-folded RV curves for the two planets orbiting HD86226 (c on the left and b on the right). Solid back curves show the best-fit Keplerian model for each planet, with light blue shading to represent the 68\%, 95\%, and 99\% posterior credibility bands. White circles show the RVs in 0.1 phase bins. Data include CORALIE98 (orange), CORALIE07 (red), CORALIE14 (purple), MIKE (blue), PFS1 (dark blue), and PFS2 (maroon). Residuals to the RV fit are shown in the bottom panels, with the same symbol representation. }
    \label{fig:joint_fit}
\end{figure*}

\begin{figure*}
    \centering
    \includegraphics[width=\textwidth]{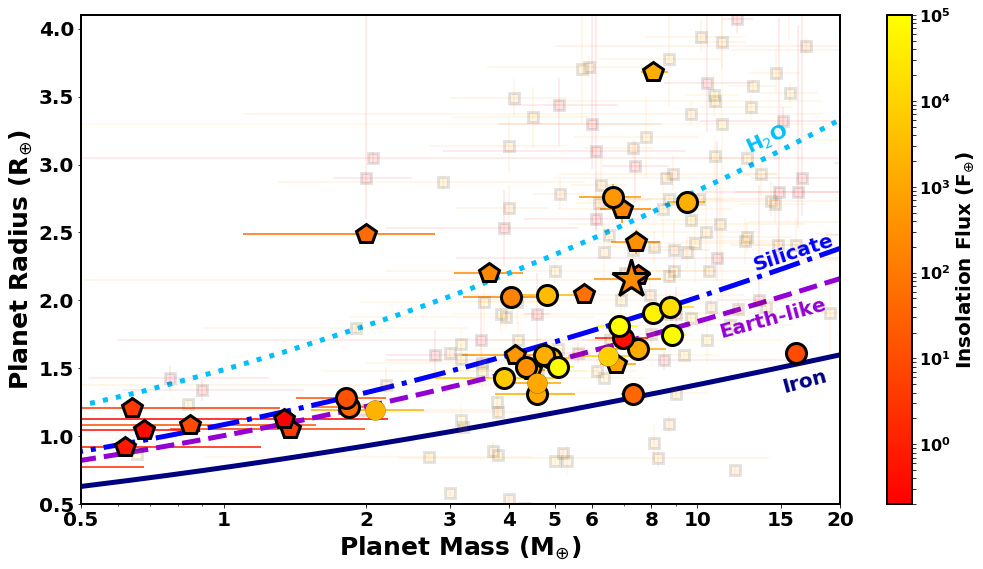}
    \caption{Visualization of where HD 86226c (bold-outlined star symbol) falls among other detected planets in terms of its mass and radius. The circle symbols correspond to planets with errors in radius and mass similar to those of HD 86226c ($\sigma_{\rm{Rp}}\leq 0.1$ $R_{\oplus}$ and $\sigma_{\rm{Mp}}\leq 1.2$ $M_{\oplus}$ errors), with masses measured via RV observations. The pentagon symbols correspond to planets with the same $\sigma_{\rm{Rp}}$ and $\sigma_{\rm{Mp}}$ limits, but with masses measured via transit-timing observations. Bold-outlined circles and pentagons correspond to multiplanet systems. The fainter background squares correspond to the remaining planets with lower-precision mass and radius measurements from either RV or TTV observations. The colored curves represent estimates of different bulk compositions -- H$_2$O (light blue dotted), silicate (dark blue dashed-dotted, no iron), Earth-like (purple dash), and pure iron (dark blue solid). See \S\ref{subsection:interior} for details on the interior modeling.}
    \label{fig:mr_interior}
\end{figure*}

The joint transit and RV fit (\S\ref{subsection:jointfit}) is shown in Figure \ref{fig:joint_fit}, and the resulting parameters are tabulated in Table \ref{tab:posteriors_joint}; all of the planet parameters have the relevant stellar parameter errors propagated. The final derived radius for HD 86226c is 2.16$\pm$0.08 $R_{\oplus}$, and the final derived mass is 7.25$^{+1.19}_{-1.12}$ $M_{\oplus}$. At a period of $\sim$4 days this planet is hot ($T_{\rm{eq}}=1311\pm28$ K), making it typical of the TESS small ($R_{\rm{p}} \leq 4 R_{\oplus}$) planets that have been detected thus far, which have a mean $T_{\rm{eq}}$ of 1270 K\footnote{From ExoFOP-TESS, \url{https://exofop.ipac.caltech.edu/tess/}, accessed on 2020 April 1}. In Figure \ref{fig:mr_interior} we show a comparison of HD 86226c and other detected planets for which the mass and radius have been measured; this figure is discussed further in \S\ref{subsection:interior}. In terms of mass and radius, HD 86226c most resembles K2-146 c (represented by a bold pentagon symbol behind the star symbol in \ref{fig:mr_interior}), Kepler-18 b, Kepler-289 b, and Kepler-48 d (represented by pale squares in the plot), but all of these other planets' host stars are $V>13$ mag. HD 86226c has the same radius and nearly to the same insolation flux as $pi$ Men c \citep[at 2.04 R$_{\oplus}$, 4.82 $M_{\oplus}$;][]{ Gandolfi2018,Huang2018}, but is 1.5$\times$ as massive.

\begin{table*}
	\centering
	\caption{Planetary Properties for HD 86226b and c}
	\begin{tabular}{lccc} 
\hline\hline        
    Property	&	HD 86226c & HD 86226b \\
	\hline\hline

    Fitted Parameters &\\ 
    	$\rho_{\star}$ (kg m$^{-3}$) & \hspace{40mm} 1233$^{+26}_{-28}$&\\
\hline

    $P$ (days)		&	3.98442$\pm$0.00018 & 1628$^{+22}_{-21}$ \\\vspace{2pt}
    $T_0$ (BJD$_{\rm TDB}$ - 2450000)&	8543.2539$\pm$0.0007 & 7308$^{+41}_{-39}$	\\\vspace{1pt}
     $a/R_{*}$		&10.11$^{+0.07}_{-0.08}$& - \\\vspace{2pt}
    $b$ & 0.63$^{+0.06}_{-0.08}$   & -\\\vspace{2pt}
    $K$ (m s$^{-1}$) 	&2.89 $^{+0.46}_{-0.43}$	& 7.74$^{+0.69}_{-0.70}$\\
    $i_{\rm p}$ (deg)  & 86.45$^{+0.26}_{-0.16}$& \\\vspace{2pt} 
     $e$ 			&0.075$^{+0.065}_{-0.048}$ & 0.059$^{+0.062}_{-0.039}$\\
    $\omega$ (deg) & 196$^{+60}_{-90}$ & 225$^{+84}_{-153}$\\\hline
    Derived Parameters & &\\
     $M_{\rm p}$ & 7.25$^{+1.19}_{-1.12}$ \me & 0.45$^{+0.04}_{-0.05}$ \mj \\\vspace{2pt}
    $R_{\rm p}$ (\re)& 2.16$\pm$0.08 & - \\\vspace{2pt}
    $a$ (au) & 0.049$\pm$0.001& 2.73$\pm$0.06\\\vspace{2pt}
     $T_D$ (hr) & 3.12$^{+0.12}_{-0.05}$  & - \\\vspace{2pt}
     $\rho_{\rm p}$ (g~cm$^{-3}$) & 3.97$^{+0.78}_{-0.73}$ & - \\\vspace{2pt}
    $T^{a}_{\rm eq}$ (K) &1311$\pm$28 & 176$\pm$4
    \\
    \hline
    \hline
    Instrumental Parameters &\\
    $M_{\mathrm{TESS}}$ (ppm)  &  0.0000068$\pm$0.0000023&\\\vspace{2pt}
    $\sigma_{w, \rm TESS}$ (ppm) &133.57$^{+4.01}_{-3.98}$&\\\vspace{2pt}
    $q_{1,\rm TESS}$ & 0.33$^{+0.29}_{-0.18}$ &\\\vspace{2pt}
    $q_{2,\rm TESS}$ & 0.38$^{+0.33}_{-0.25}$&\\
    \hline
    \underline{Instrument} & \underline{$\mu$ (m s$^{-1}$)} & \underline{$\sigma_{w}^{b}$ (m s$^{-1}$)} & \underline{$N_{\rm obs}$}\\ 
    COR98 & 19744.07$^{+2.59}_{-2.72}$ & 7.87$^{+2.73}_{-1.94}$ & 12 \\
    COR07 & 19739.44$^{+1.12}_{-1.18}$ & 7.22$^{+1.01}_{-0.88}$ & 50  \\ 
    COR14 & 19761.52$^{+1.29}_{-1.32}$ & 5.06$^{+1.17}_{-1.03}$ & 24 \\
    MIKE  & -4.57$^{+1.15}_{-1.21}$    & 4.21$^{+2.89}_{-2.54}$ & 13 \\
    PFS1  & -4.41$^{+0.63}_{-0.60}$    & 4.27$^{+0.58}_{-0.49}$ & 44 \\
    PFS2  & -6.43$^{+0.78}_{-0.79}$    & 1.70$^{+0.35}_{-0.30}$ & 23 \\
    \hline\hline
    \multicolumn{3}{l}{$^{a}$ Equilibrium temperature estimated considering Bond Albedo $A_{B}=0$.}\\
    \multicolumn{3}{l}{$^{b}$ Instrumental jitter term added in quadrature to measurement errors listed in }\\
    \multicolumn{3}{l}{Tables \ref{tab:COR} and \ref{tab:pfsrvs} to produce the error bars shown in Figure \ref{fig:joint_fit}, bottom panel.}\\

	\end{tabular}
    \label{tab:posteriors_joint}
\end{table*}

        \begin{figure}
            \centering
            \includegraphics[width=\columnwidth]{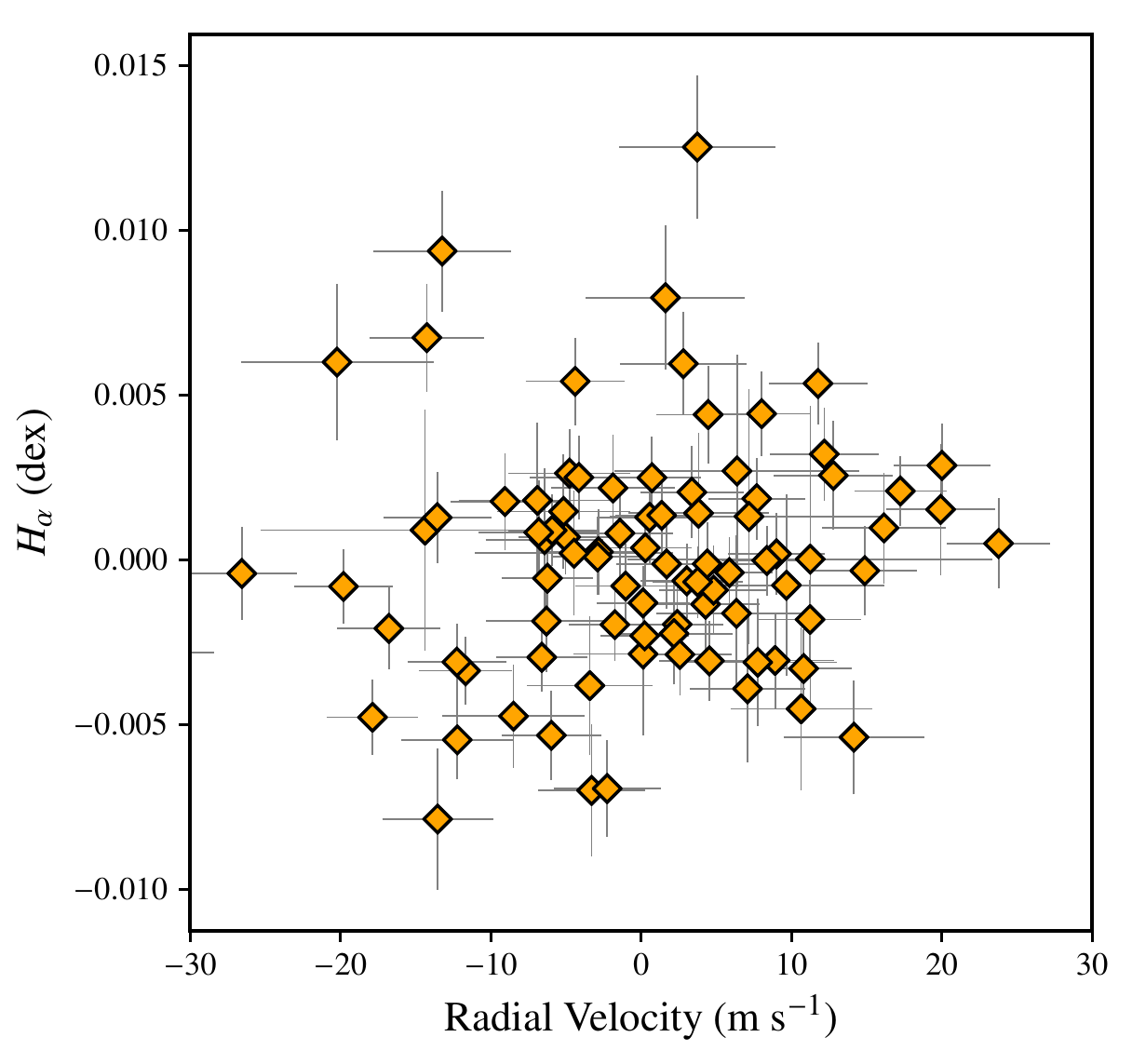}
            \includegraphics[width=\columnwidth]{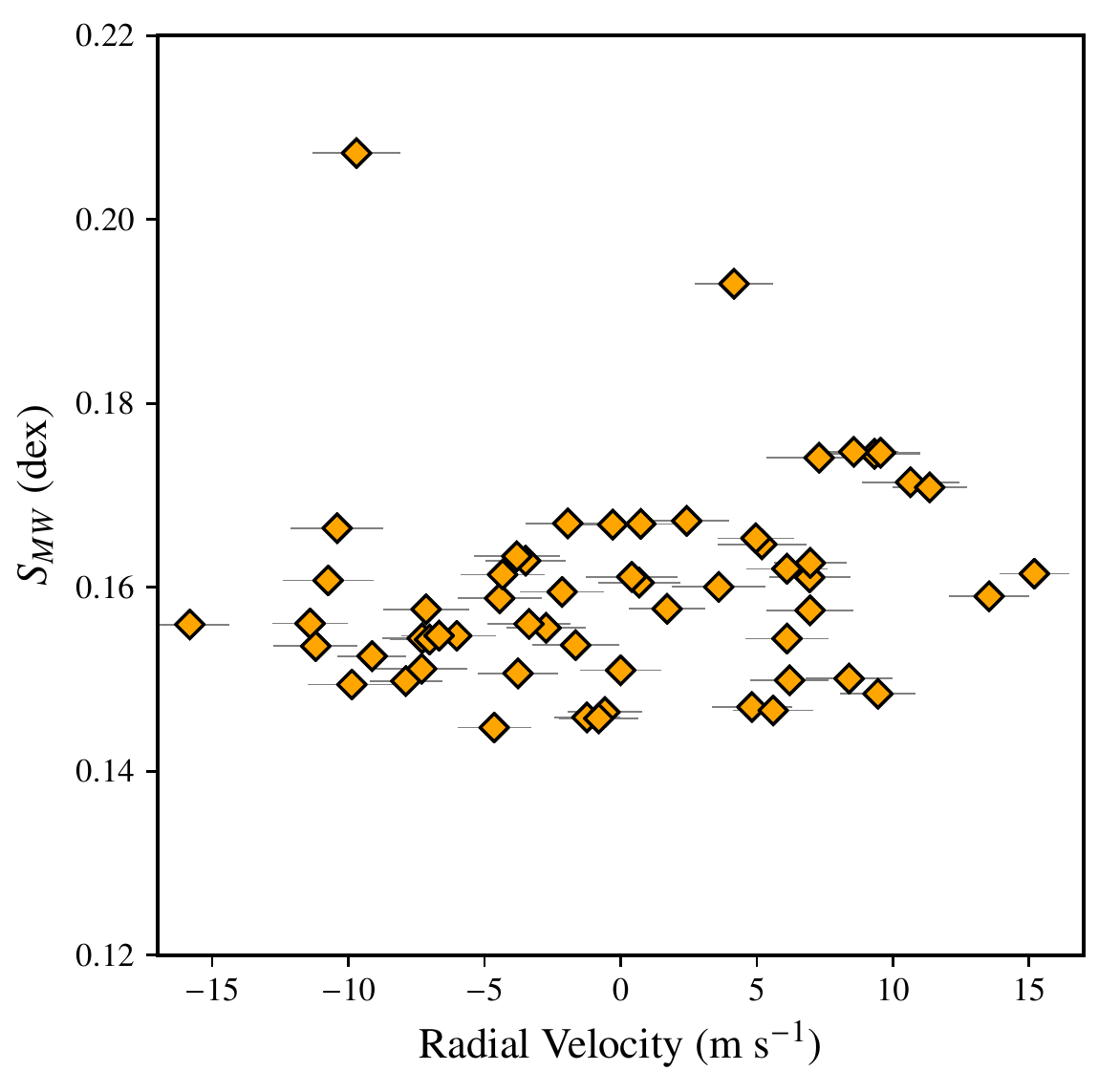}
            \caption{Top: Correlations between RVs and $H_{\alpha}$ index from CORALIE. Bottom: Correlations between RVs and S-index ($S_{\rm MW}$) values from PFS. In each plot we have subtracted the mean value of both RV and the activity indices. 
            }
            \label{fig:rv_corr}
        \end{figure}

    \subsection{Stellar Variability \label{subsec:stellar_variability}}
    Our analysis of the photometry in \S\ref{sec:new_obs} does not give a clear measurement of the rotation period of HD 86226, which is perhaps to be expected given only one sector of \tess{} data ($\sim$28 days). From the TESS photometry, the best-fit variability period is 6.4$\pm$0.7 days, but with a low amplitude of 79$\pm$3 ppm. Given the lower precision of the ground-based photometric measurements, we do not see this variability in the WASP-South data, nor do we see peaks at such a short period in the ASAS or ASAS-SN GLS periodograms (Figures \ref{fig:asas_per} and \ref{fig:asassn_per}); these data seem to indicate the most power at $\gtrsim 80$ days. The S-index and H$\alpha$ variability time series also show the most significant power at longer periods, although with a larger number of less distinct peaks than are present in the ASAS or ASAS-SN photometry. In any case, given the resemblance of HD 86226 to the Sun in temperature and age (see Table \ref{tab:exofast}), it is unlikely that its rotation period is as short as six days or the planet orbital period of four days \citep[e.g., ][]{McQuillan2014}. 
    
    Recent work by \cite{Nava2019} shows that, due to the uneven and evolving nature of magnetic active regions in the atmospheres of stars, signals related to magnetic activity can cause significant peaks in the RV periodograms that do not correspond to the stellar rotation period or even its harmonics. This is true even when the active region lifetime is much greater than the rotation period, such that one would expect the rotation signal to dominate. The authors caution that spurious periodogram peaks are inherent in RVs across many different distributions of stellar activity, such that spurious power could be added at planet periods and thus contributed to inaccurate mass determinations. Detailed exploration of different magnetic activity models, similar to what Nava et al. did for K2-131 and Kepler-20, would provide an additional test on the robustness of our derived masses, but is beyond the scope of this paper.

    We also checked for linear correlations between the H$\alpha$ activity and CORALIE RV measurements (Figure \ref{fig:rv_corr}, top) and the S-index activity and PFS RV measurements (Figure \ref{fig:rv_corr}, bottom). While there are slight positive correlations in both cases,they are not significant as determined by the Pearson (testing a linear relationship) and Spearman rank (testing a monotonic relationship) correlation coefficients, which are $r = 0.10\pm0.4$ and $\rho=0.13$  ($p$-value = 0.24, sample size = 88) for CORALIE, and $r = 0.18\pm0.2$ and $\rho=0.3$ ($p$-value = 0.02, sample size = 58) for PFS, respectively. Here, the errors on the $r$ values were determined via jackknife resampling.

\section{Discussion and Conclusions}

    \subsection{Interior Composition Estimates for HD 86226c}\label{subsection:interior}

    To explore the range of possible compositions of HD 86226c, we modeled the interior considering a pure iron core, a silicate mantle, a pure water layer, and a H-He atmosphere. The thickness of the planetary layers were set by defining their masses and solving the structure equations. To obtain the transit radius, we follow \cite{Guillot2010} and evaluate where the chord optical depth $\tau _{\rm{ch}}$ is $2/3$. We followed the thermodynamic model of \cite{Dorn2017}, with the equation of state (EOS) of the iron core taken from \cite{Hakim2018}; the EOS of the silicate mantle is calculated with PERPLE\_X from \cite{Connolly09} using the thermodynamic data of \cite{stixrude_thermodynamics_2011}, and the EOS for the H-He envelope is calculated assuming protosolar composition, based on the semi-analytical H/He model of Saumon \& Chabrier  \cite[e.g.][SCvH]{Saumon1995}. For water, we used the QEOS of \cite{Vazan13} for low pressures and that of \cite{Seager17} for pressures above 44.3 GPa. Our input values for the model were the derived planet mass, radius, and the stellar abundances from \cite{maldonado2018}. Figure \ref{fig:mr_interior} shows $M-R$ curves tracing compositions of pure iron, Earth-like, silicate, and pure water. The silicate composition line is computed with the oxides Na$_2$O-CaO-FeO-MgO-Al$_2$O$_3$-SiO$_2$, and the Mg/Si and Fe/Si ratios of the Earth's mantle. The water line corresponds to a surface pressure of 1\,bar, which corresponds to water worlds without water vapor atmospheres. 
    As shown in Figure \ref{fig:mr_interior}, there are several small planets following the Earth-like composition with relatively small dispersion up to about 7 $M_{\oplus}$. HD 86226c lies slightly above the silicate composition line, suggesting that it is richer in water or volatile elements like H-He than most of the planets below 7 $M_{\oplus}$ detected so far. Therefore, it might represent a new type of terrestrial planet that differs significantly from Earth in terms of bulk composition.

    We then used a generalized Bayesian inference analysis with a nested sampling scheme \cite[e.g.][]{Buchner2014} to quantify the degeneracy between interior parameters and produce posterior probability distributions. The interior parameters that were inferred include the masses of the pure iron core, silicate mantle, water layer and H-He atmospheres. We assumed the Fe/Si and Mg/Si ratios inside the planet are the same as the ratios observed in the stellar photosphere: 0.79 and 1.15, respectively, from \citealt{maldonado2018}. Table \ref{tab:interior} lists the inferred mass fractions of the core, mantle, and water layer and H-He atmosphere from our structure models. We found that HD 86226c has a H-He envelope of 4.6$\times 10^{-4} M_{\oplus}$ and thickness of 0.39$R_{\oplus}$. The other three constituents of the planet have relative mass fractions between 32\% and 35\% with large uncertainties (see Table \ref{tab:interior}). This regime of the $M-R$ relation is strongly degenerate, and therefore even with more precise mass and/or radius measurements, it would not be possible to significantly improve the estimate of the mass ratio between the core, mantle, and water layer. Atmospheric characterization will be crucial to better constrain the volatile envelope ratio and composition.

\begin{table}
\centering
\caption{Inferred interior structure properties of HD 86226c. \label{tab:interior}}
\begin{tabularx}{9.cm}{X}
\hline
\\
    $M_{\rm{core}}/M_{\rm{total}}$  \ \ \     ..............................  0.35 $^{+0.22} _{-0.16}$\\

   $M_{\rm{mantle}}/M_{\rm{total}}$ ..............................  0.33 $^{+0.24} _{-0.20}$\\
   $M_{\rm{water}}/M_{\rm{total}}$ \ ..............................  0.32 $^{+0.21} _{-0.17}$\\
   $M_{\rm{atm}}/M_{\rm{total}}$ \ \ \  ..............................  0.000062 $^{+0.000027} _{-0.000023}$\\
\\
  \hline
\end{tabularx}
\end{table}

    \subsection{Potential for Atmospheric Characterization of HD 86226c}\label{sec:atm}

     \begin{figure}[thb!]
     \centering
     \resizebox{\columnwidth}{!}{\includegraphics[trim=2.0cm 9.4cm 2.0cm 10.0cm, clip]{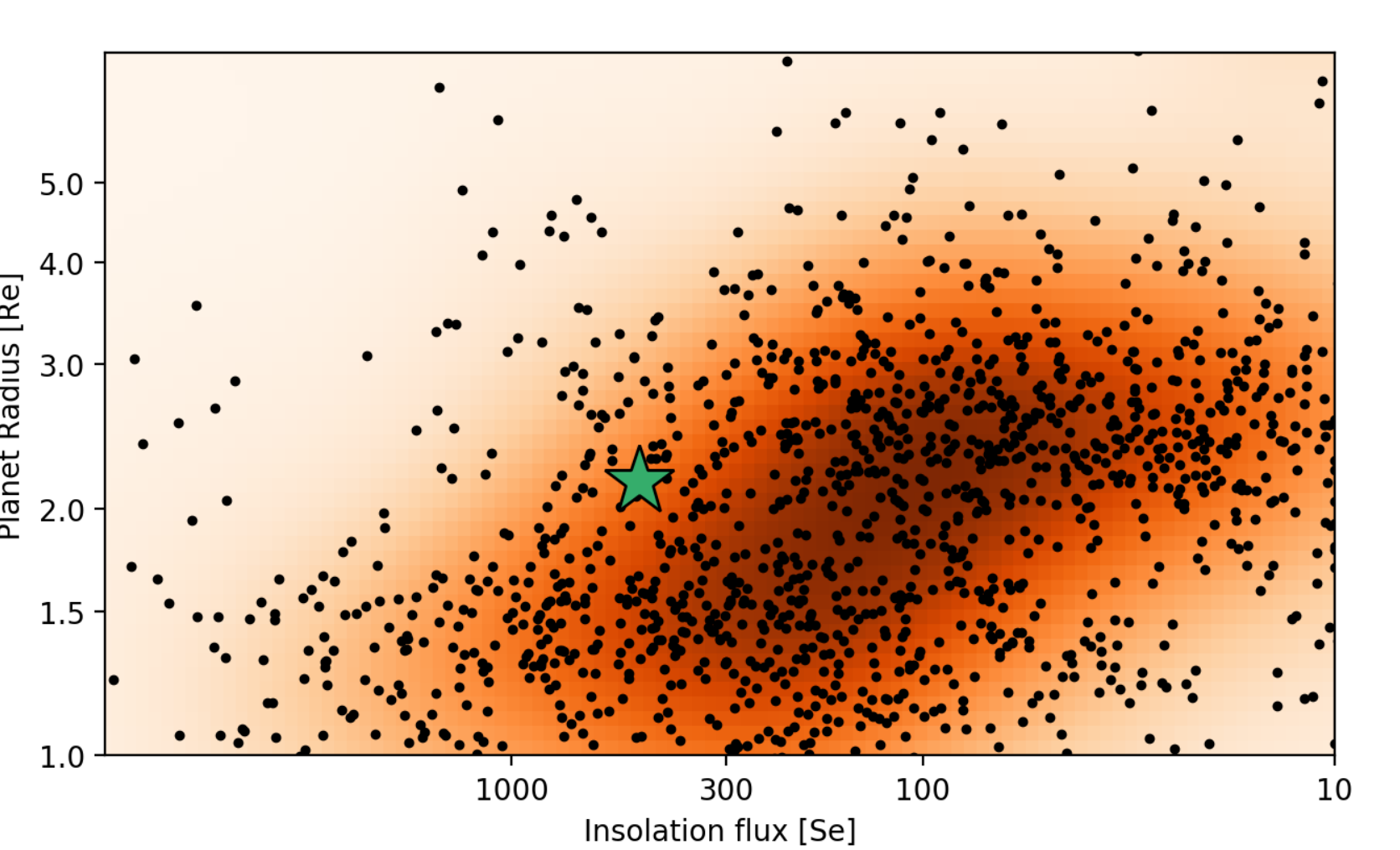}}
    \caption{\label{fig:insolation} Insolation flux relative to Earth as a function of planet radius, extracted from the NASA Exoplanet Archive. The data are presented in similar fashion to \cite{Fulton2017, Fulton2018}. The shading represents point density, showing the different populations of sub-Neptunes and super-Earths, with dense populations shown in darker shades. The light area in the upper left of the plot indicates the hot Neptune desert. HD 86226c is shown as a green star towards the edge of the desert.}
    \end{figure}

    Figure \ref{fig:insolation} shows the position of HD 86226c in the radius-insolation space. The planet is located toward the edge of the hot Neptune desert \citep{Szabo&Kiss2011,Beauge&Nesvorny2013,Mazeh2016}, making it an interesting target to study the processes leading to this dearth of planets \citep[e.g.,][]{Brunini&Cionco2005, Helled2016, Owen&Lai2018}. It is not located within the desert, however, which makes it more likely that the planet retained its atmosphere despite the intense radiation from the star \citep{Owen2012,Lopez2013,Owen2013,Jin2014,Lopez2017}. The detection of an escaping envelope, or a stringent upper limit, could add a valuable data point along the edge of this desert. To date, atmospheric studies have only been conducted for
    a small number of exoplanets in this regime, such as GJ436b \citep{Butler2004}, GJ3470b \citep{Bonfils2012}, and HAT-P-11b \citep{Bakos2010}.

    The escaping atmospheres of exoplanets have been previously detected using the cross-correlation function technique and high-resolution spectroscopic data obtained during transit \citep{Hoeijmakers2018, Nortmann2018}.The escaping envelope could be observed via the He I triplet in the infrared \citep{Seager2000, Turner2016, Oklopcic2018, Oklopcic2019} and the Balmer series of H I in the visible range \citep{Jensen2012, Cauley2017, Jensen2018, Yan2018, Casasayas2019}. Additionally, if sodium exists in its neutral state in the exoplanet atmosphere, it can be a tracer for the upper layers up to the thermosphere of the planet \citep{Redfield2008}. The size of the planet (transit depth of $\sim$400 ppm in the TESS bandpass) makes the use of a 
  large aperture telescope necessary 
  for such high-resolution, ground-based observations of its atmosphere.

    HD 86226c is observable from the Southern Hemisphere and is thus a prime target for ESPRESSO at the VLT.\footnote{Although HD 86226 is visible from Maunakea Observatory for a few hours between mid-January and mid-March, there are no transits of planet c observable in those windows in the next year.} ESPRESSO's wavelength range covers both the Na and H$\alpha$ spectral features. Using ESO's tools to calculate necessary exposure times for a signal to noise of 100 per exposure, one could achieve up to 29 exposures during one transit\footnote{\footnotesize{Calculated with ESO's Astronomical Tools \url{http://eso.org/sci/facilities/paranal/sciops/tools.html}.}}. Comparing with similar observations performed with ESPRESSO \citep{Chen2020}, we estimate that the sodium feature, if it exists, can be detected at the $5\sigma$ level by combining three transits. Additionally, metals (e.g., Mg, Ti, Fe) could also be searched for in the atmosphere of HD 86226c, thanks to ESPRESSO's blue wavelength coverage. 
    Searching for near-infrared water features in the atmosphere of HD 86226c is also feasible right now with CRIRES+ at the VLT \citep{Follert2014}, and will be in the future with the HIRES optical-to-NIR spectrograph at the E-ELT \citep{Marconi2016}.

    Tracing the potentially escaping volatile envelope of HD 86226c is also possible via the Lyman-$\alpha$ line. Due to the close proximity ($45.57$~pc) and the solar-type host star, any potential Lyman-$\alpha$ signal should be detectable with space-borne spectrographs despite absorption by the interstellar medium, as has been detected for similarly sized planets in the past, most notably GJ436~b \citep{Kulow2014, Ehrenreich2015, Bourrier2015, Bourrier2016, Lavie2017}. The planet's size makes the study of these lines challenging with the Hubble Space Telescope (detection limits were simulated with the Pandexo Exposure Time Calculator\footnote{\footnotesize{ \url{https://exoctk.stsci.edu/pandexo/}.}}). However, HD 86226c is a potential target for the James Webb Space Telescope in the future, not only to trace the upper atmosphere but also to search for a potential water feature in the infrared with NIRSpec.

    \subsection{HD 86226c: Another Small Planet with a Jupiter Analog}

As discussed in the introduction, there is growing evidence that some percentage ($\sim$30-40\%) of stars hosting small planets also host a larger, longer-period planet. This observation is intriguing because it suggests that the presence of outer gas giant planets does not hinder the formation of inner smaller planets, and in fact may facilitate the growth of some subset of small planets. Whether there are differences in the properties of small planets that have or do not have giant planet companions is thus an interesting --  but still open -- question. 

        \begin{figure}[t]
            \centering
            \includegraphics[width=\columnwidth]{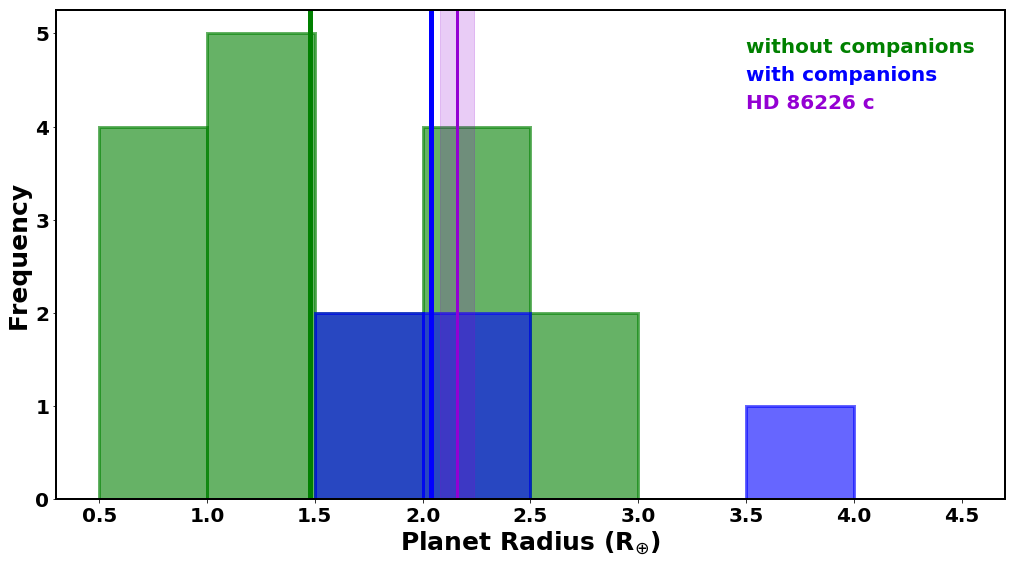}
            \includegraphics[width=\columnwidth]{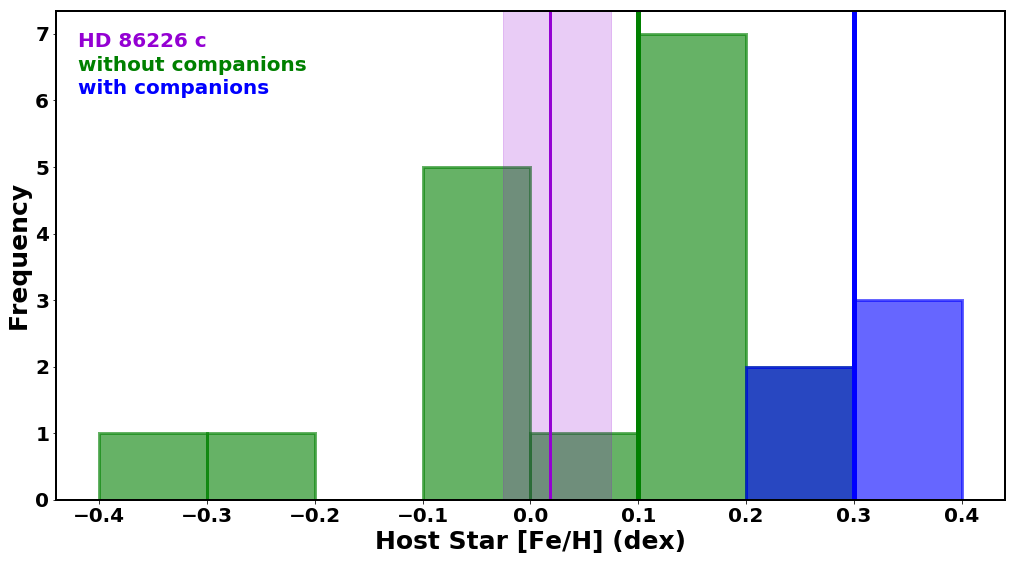}
            \caption{Here we show a comparison of HD 86226c (purple, with shading corresponding to 1$\sigma$ errors) to planets within $\pm2\sigma$ of its mass from the B19 sample. Top: The green histogram (vertical line) represents the distribution 
            (median) of the radii of planets without giant planet companions. The blue histogram (vertical line) represents the distribution (median) of the radii of planets with giant planet companions. Bottom: The green histogram (vertical line) represents the distribution (median) of the host star metallicities of planets without giant planet companions. The blue histogram (vertical line) represents the distribution (median) of the host star metallicities of planets with giant planet companions.}
            \label{fig:bryan_compare}
        \end{figure}

To place HD 86226c in context, we compared its properties to those of the planets in the \citet[][hereafter B19]{bryan2019} sample, which consists of systems with at least one confirmed planet with either a mass between 1 and 10 $M_{\oplus}$ or a radius between 1 and 4 $R_{\oplus}$, depending on the detection technique. Each of the 65 systems in the B19 sample have at least 10 published RV data points across a baseline of at least 100 days, allowing the authors to search for long-period giant companions (with either resolved orbits or statistically significant linear trends). However, we note that the sensitivity to long-period companions is different, on average, for those systems with the inner planet detected via RV versus transit. For example, B19's data and analysis would typically be sensitive to a 1 $M_{\rm{Jup}}$ planet at about 6 AU in the RV case, but only out to about 1.5 AU for the typical transit case. The RV-detected systems have greater sensitivity because they typically have a longer baseline of RV data (see Figure 5 in B19), therefore the sample of small planets detected via transit that have giant planet companions could be artificially small. Given the existing RV data on HD 86226, it would be included in the RV sample of B19.

When we restricted the B19 sample to only planets with measured masses within $\pm2\sigma$ of HD 86226c's mass there were 22 planets that also had measured radii, 17 that did not show evidence of a giant planet companion (``without companion'' sample) and five that did (``companion'' sample). In the top panel of Figure \ref{fig:bryan_compare} we show how HD 86226c compares in radius to these 22 similar-mass planets from B19. HD 86226c's radius is just above the median radius of the ``with companions'' sample ($\widetilde{R_{p}}=$2.04 $R_{\oplus}$; blue vertical line). Interestingly, the top panel of Figure \ref{fig:bryan_compare} may point toward a potential difference in size between small planets with and without giant planet companions -- it appears that small planets \textit{without} giant planet companions (green distribution, $\widetilde{R_{p}}=$1.48 $R_{\oplus}$) may generally extend to smaller radii. 

In the bottom panel of Figure \ref{fig:bryan_compare} we show a comparison of host star metallicities ([Fe/H]; taken from B19). Planets with masses similar to that of HD 86226c that have a giant planet companion (in blue) are skewed toward higher host star [Fe/H] ($\widetilde{[Fe/H]}=$0.30 dex) versus those without giant planet companions (in green, $\widetilde{[Fe/H]}=$0.10 dex). B19 also found evidence in their full sample that planets without giant companions orbited lower-metallicity host stars than planets with giant companions. If host star [Fe/H] is a proxy for protoplanetary disk solid material content, then perhaps small planets are larger in systems with giant planet companions because there was more material around to form the core faster and thus more readily acquire an atmosphere \citep{dawson2015,owen&murray-clay2018}. This is in contrast to \cite{Buchhave2018} who found that stars hosting Jupiter analogs (1.5-5.5 AU, $e<0.25$, 0.3-3 $M_{\rm{J}}$ planets) are on average closer to solar metallicity ([Fe/H] = 0). Based on numerical simulations, Buchhave et al. suggested that metal-rich systems form multiple Jupiter analog planets, leading to planet-planet interactions and therefore eccentric cool or circularized hot Jupiter planets. In the case of HD 86226c, Figure \ref{fig:bryan_compare} also shows that its host star [Fe/H] is below both the ``with companion'' and ``without companion'' median values -- even though HD 86226c has a long-period giant planet companion, its host star [Fe/H] is $\sim$0. Again, we note that some of the ``without companion'' sample may actually contain small planets with companions, if they remain undetected due to lack of RV data. Perhaps this contributes to the difference between the metallicity of HD 86226 and that of host stars of other similar-mass small planets with companions.

\begin{figure}[h]
            \centering
            \includegraphics[width=\columnwidth]{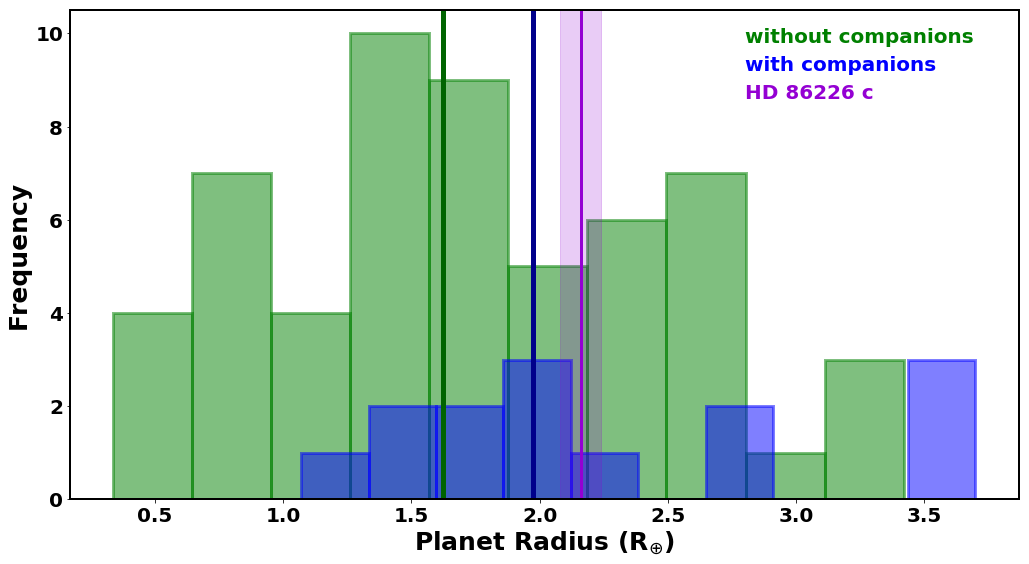}
            \includegraphics[width=\columnwidth]{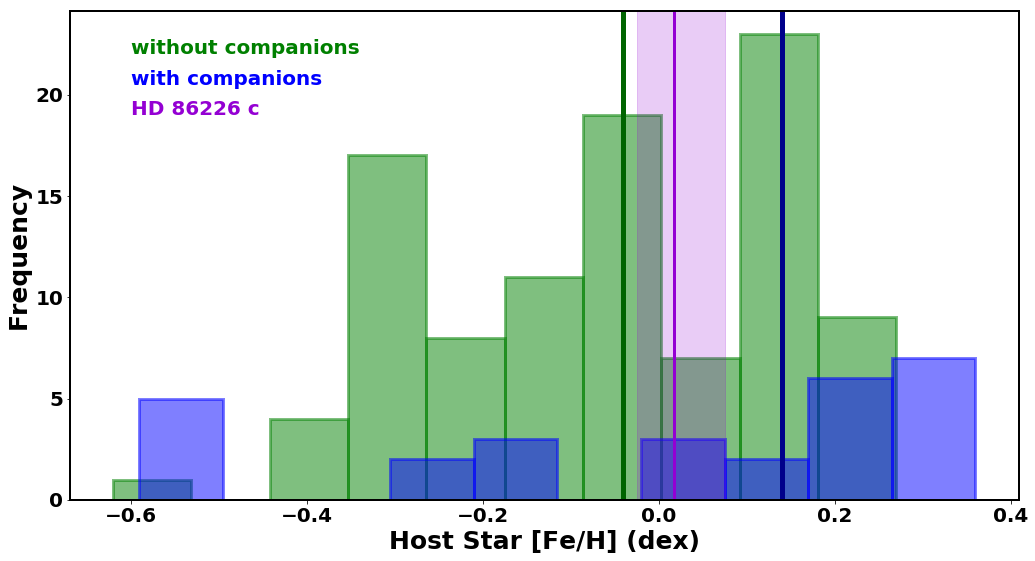}
            \caption{Similar to Figure \ref{fig:bryan_compare}, only here we show a comparison of HD 86226c to the full B19 sample.}
            \label{fig:bryan_compare_full}
        \end{figure}

Thus far, we have only compared HD 86226c to the B19 planets having measured masses within $\pm2\sigma$ of HD 86226c. Given this small subset, we wanted to see if the same trends with radius and host star metallicity held in the full sample. In Figure \ref{fig:bryan_compare_full}, top panel, HD 86226c is again slightly above the median radius of the ``with companion'' sample. In this figure, we see that full sample of radii of planets without companions (green) tends to be shifted toward radii slightly smaller ($\widetilde{R_{p}}=$1.62 $R_{\oplus}$) than those of planet with companions (blue; $\widetilde{R_{p}}=$1.98 $R_{\oplus}$). A comparison of the B19 planets with radii $<$4 $R_{\oplus}$ to the full sample of known planets with $<$4 $R_{\oplus}$\footnote{Retrieved from the NASA Exoplanet Archive on 2020 February 27.} showed no significant difference in their radii distributions, so we do not expect this result to be heavily biased by the B19 sample selection. 

In the bottom panel of Figure \ref{fig:bryan_compare_full} we plot the full B19 sample of host star metallicities. Here, the means of the ``no companion'' (green, $\widetilde{[Fe/H]}=$-0.04 dex) and ``with companion'' (blue, $\widetilde{[Fe/H]}=$0.14 dex) are shifted to lower values but have roughly the same offset as the mass-restricted sample above. The metallicity of HD 86226 falls in between ``with'' and ``without'' companion host star median [Fe/H] values. Again, a comparison of the B19 planets with radii $<$4 $R_{\oplus}$ to the full sample of known planets with $<$4 $R_{\oplus}$ showed no significant difference in their host star metallicity distributions, so the B19 sample is not biased toward higher or lower host star metallicities. 

In the analysis above, there are hints that small planets without giant planet siblings typically have a wider radius distribution, extending to smaller radii. A full statistical treatment including  potential biases in the sample is beyond the scope of this work. However, it is interesting to consider the following questions: \textit{if} small planets in systems with giant planets are systematically larger, what is the origin of this difference? Is it just an effect of higher metallicity enabling the formation of bigger planets, as suggested above? Or, perhaps there is something specifically related to the formation and evolution of a giant planet in the system that makes the small planets larger? A survey searching for giant planet companions to small planets with host stars in a limited [Fe/H] range would isolate one effect (host star [Fe/H]) while letting the other (presence of giant planet companion) vary, and thus could potentially help address the origin of the size difference in small planets. In such a survey, it would also be important to consider the potential effects of the host star irradiation over time on the size of the small planets. Given its low host star metallicity, size, and presence of an outer companion, HD 86226c may be important in addressing potential differences in the population of small planets with giant companions versus those without. 

\acknowledgments

Funding for the TESS mission is provided by NASA's Science Mission directorate. We acknowledge the use of public TESS Alert data from pipelines at the TESS Science Office and at the TESS Science Processing Operations Center. Resources supporting this work were provided by the NASA High-End Computing (HEC) Program through the NASA Advanced Supercomputing (NAS) Division at Ames Research Center for the production of the SPOC data products. This research has made use of the Exoplanet Follow-up Observation Program website, which is operated by the California Institute of Technology, under contract with the National Aeronautics and Space Administration under the Exoplanet Exploration Program. This paper includes data collected by the TESS mission, which are publicly available from the Mikulski Archive for Space Telescopes (MAST). This work makes use of observations from the LCOGT network. Support for this work was provided by NASA through Hubble Fellowship grant HST-HF2-51399.001 awarded by the Space Telescope Science Institute, which is operated by the Association of Universities for Research in Astronomy, Inc., for NASA, under contract NAS5-26555. Support for this work was also provided by NASA through grant 18-XRP18\_2-0048. M.R.D. acknowledges the support of CONICYT-PFCHA/Doctorado Nacional-21140646, Chile. R.L. has received funding from the European Union's Horizon 2020 research and innovation program under the Marie Sklodowska-Curie grant agreement No.~713673 and financial support through the "la Caixa" INPhINIT Fellowship Grant LCF/BQ/IN17/11620033. JVS is supported by funding from the European Research Council (ERC) under the European Union's Horizon 2020 research and innovation programme (project {\sc Four Aces}; grant agreement No. 724427). J.S.J. acknowledges support by FONDECYT grant 1161218 and partial support from CONICYT project Basal AFB-170002. E.P. acknowledges funding from the Spanish Ministry of Economics and Competitiveness through project PGC2018-098153-B-C31. J.N.W. thanks the Heising-Simons Foundation for support. This work has been carried out within the framework of the National Centre of Competence in Research PlanetS supported by the Swiss National Science Foundation. J.H. acknowledges the support from the Swiss National Science Foundation under grant 200020\_172746. Part of this research was carried out at the Jet Propulsion Laboratory, California Institute of Technology, under a contract with the National Aeronautics and Space Administration (NASA). We thank George Zhou for his time observing with Magellan II/PFS, Dan Foreman-Mackey for his invaluable help with custom configurations of \texttt{exoplanet}, and the anonymous referee for their helpful comments that improved the quality of this work.

\facilities{\tess{}, Magellan:Clay(PFS), \\
Euler1.2m(CORALIE), LCOGT}
\software{astropy \citep{astropy}, emcee \citep{emcee}, Lightkurve \citep{Lightkurve}, AstroImageJ \citep{Collins:2017}}
\bibliographystyle{aasjournal}
\bibliography{references}


\end{document}